\documentclass[useAMS,usenatbib]{mn2e}

\usepackage{verbatim}
\usepackage[english]{babel}
\usepackage{graphicx}
\usepackage{color}
\usepackage{amssymb}
\usepackage{amsfonts}
\usepackage{amsmath}

\DeclareOldFontCommand{\sc}{\normalfont\scshape}{\@nomath\sc}

\newcommand{\hi}{H\,{\sc i}}
\newcommand{\hii}{H\,{\sc ii}}

\newcommand{\he}{He}
\newcommand{\hei}{He\,{\sc i}}
\newcommand{\heii}{He\,{\sc ii}}
\newcommand{\heiii}{He\,{\sc iii}}

\newcommand{\mh}{$h^{-1}\textrm{Mpc}$}
\newcommand{\kh}{$h^{-1}\textrm{kpc}$}

\newcommand{\Myr}{$\textrm{Myr}$}
\newcommand{\kms}{$\tr{km}\,\tr{s}^{-1}$}

\newcommand{\sunh}{$h^{-1}\textrm{M}_\textrm{\sun}$}
\newcommand{\tr}{\textrm}
\newcommand{\ti}{\textit}
\newcommand{\be}{\begin{equation}}
\newcommand{\ee}{\end{equation}}
\newcommand{\bea}{\begin{eqnarray}}
\newcommand{\eea}{\end{eqnarray}}

\newcommand{\tn}{\kern 0.16667 em}

\newcommand{\aj}{AJ}
\newcommand{\apj}{ApJ}
\newcommand{\apjl}{ApJL}
\newcommand{\apjs}{ApJS}
\newcommand{\mnras}{MNRAS}
\newcommand{\aap}{A\&A}
\newcommand{\araa}{ARA\&A}
\newcommand{\nat}{Nature}

\title[The imprint of \heii\ reionization]{The imprint of inhomogeneous \heii\ reionization on the \hi\ and \heii\ Ly$\alpha$ forest}

\author[M. Compostella, S. Cantalupo \& C. Porciani]{Michele Compostella$^{1}$\thanks{E-mail:\newline
mcompos@astro.uni-bonn.de (MC);\newline cantal@ucolick.org (SC);\newline porciani@astro.uni-bonn.de (CP).}, Sebastiano Cantalupo$^{2,3}$\footnotemark[1] and Cristiano Porciani$^{1}$\footnotemark[1]
\\
$^{1}$Argelander Institut f\"{u}r Astronomie der Universit\"{a}t Bonn, Auf dem H\"{u}gel 71, Bonn, D-53121, DE.
\\
$^{2}$Department of Astronomy and Astrophysics, UCO/Lick Observatory, University of California, 1156 High
Street,\\ Santa Cruz, CA 95064, USA.
\\
$^{3}$Kavli Institute for Cosmology, Cambridge \& Institute of Astronomy, Madingley Road, Cambridge CB3 0HA, UK.
}

\begin{document}

\date{Accepted ; Received ; in original form .}

\pagerange{\pageref{firstpage}--\pageref{lastpage}} \pubyear{2013}

\maketitle

\label{firstpage}

\begin{abstract}
We use a set of adaptive mesh refinement hydrodynamic simulations post-processed with the radiative-transfer code {\sevensize RADAMESH} to study how inhomogeneous \heii\ reionization affects the intergalactic medium (IGM). We propagate radiation from active galactic nuclei (AGNs) characterized by a finite lifetime and located within massive dark matter haloes. We consider two scenarios for the time evolution of the ionizing sources. In all cases,  we find that \heii\ reionization takes place in a very inhomogeneous fashion, through the production of well-separated bubbles of the ionized phase that subsequently percolate. Overall, the reionization process is extended in time and lasts for a redshift interval $\Delta z \gtrsim 1$.
At fixed gas density, the temperature distribution is bimodal during the early phases of \heii\ reionization and cannot be described by a simple effective equation of state.
When \heii\ reionization is complete, the IGM is characterized by a polytropic equation of state with index $\gamma \simeq 1.20$. This relation is appreciably flatter than at the onset of the reionization process ($\gamma=1.56$) and also presents a much wider dispersion around the mean. We extract \hi\ and \heii\ Ly$\alpha$ absorption spectra from the simulations and fit Voigt profiles to them. Our results are fully consistent with the observed evolution in the mean transmission and the flux power spectrum. We find that the regions where helium is doubly ionized are characterized by different probability density functions of the curvature and of the Doppler $b$ parameters of the \hi\ Ly$\alpha$ forest as a consequence of the bimodal temperature distribution during the early phases of \heii\ reionization.
The column-density ratio in \heii\ and \hi\ shows a strong spatial variability. Its probability density function rapidly evolves with time reflecting the increasing volume fraction in which ionizing radiation is harder due to the AGN contribution.
Finally, we show that the number density of the flux-transmission windows per unit redshift and the mean size of the dark gaps in the \heii\ spectra have the potential to distinguish between different reionization scenarios.
\end{abstract}


\section[]{INTRODUCTION}

Observations of the rest-frame ultraviolet (UV) spectra of quasars at redshift $z>2$ show
that hydrogen and helium atoms in the intergalactic medium (IGM) have been
nearly fully ionized within the first few Gyr of
cosmic history \citep[see][for a detailed review]{Fan, Meiksin2009}.
The epoch witnessing the transition from a primarily neutral to an ionized
IGM is generally called cosmic reionization.
Photoionization due to sources of hard electromagnetic radiation is believed
to be the physical process responsible for it.
This is the most dramatic manifestation of the feedback occurring between
luminous sources and the IGM.

Given the increasing ionization potential of
\hi, \hei\ and \heii, the reionization of these species may have occurred at
different epochs, depending on the density, luminosity
and spectral hardness of the photon sources.
Uncovering when and how these processes
took place can therefore provide key information for both the study of the
high-redshift
IGM and the luminous component of the Universe, e.g. galaxies and active
galactic nuclei (AGNs).

The reionization of \hi\ was likely completed within the redshift interval
$6<z<12$,
as evidenced by the transmission of Ly$\alpha$ photons in the spectra of
$z\sim6$
quasars \citep[e.g.][]{Fan_et_al} and by the polarization of the
cosmic microwave background on large angular scales \citep[e.g.][]{Komatsu}.
During this epoch,  \hi\ and \hei\ ionizing photons were likely provided by the
abundant Population II stars associated with the early stages of galaxy
formation \citep[e.g.][]{Sokasian}.
These sources, however, are much less efficient to doubly ionize helium, given
the substantial lack
of photons at energies above $4\,\tr{ryd}$ in their spectra. It is thus believed
that \heii\ reionization is delayed to later epochs, when sources with harder
spectra, like quasars and other AGNs,
become more readily available \citep[see e.g.][]{HM2012}.

Several lines of evidence based on astronomical observations support this
picture.
The Gunn--Peterson test applied
to the quasar Ly$\alpha$ forest  --
the most direct way to probe the final stages of both \hi\ and \heii\
reionization --
shows a larger opacity in \heii\ than in \hi\ at $z\sim3$
\citep[e.g.][]{Syphers_S, Shull, Worseck}.
Unfortunately, this comparison is difficult because of the challenges to
directly observe the \heii\ Ly$\alpha$ forest of high-$z$ quasars.
In fact, \heii\ studies
are restricted to space-based observations
(given the \heii\ Ly$\alpha$ rest-frame
wavelength of $304\,\mathrm{\AA}$) and also by the paucity of bright quasars
for which the far-UV is not extinguished by intervening \hi\ absorption
(the so called `\heii\ quasars').
Cross-matching quasar catalogues with space-based UV imaging has recently
increased by a large factor the number of \heii\ quasars
\citep[e.g.][]{Syphers, Worseck_Prochaska, Worseck, Syphers2012}.
However, the number of \heii\ quasars bright enough for high signal-to-noise ratio
studies remains very limited, consisting of a handful of sources to date
\citep[see e.g.][]{Jakobsen1994, Davidsen1996, Hogan1997, Reimers1997, Zheng2004, Reimers2005, Fechner2006}.
Observations of these quasars suggest a patchy reionization history with
high \heii\ optical depths, on average, at $z>3$ and
a slow recovery of transmitted flux at $z<2.7$ \citep[e.g.][]{Shull, Worseck}.
A large scatter between sightlines and the presence of several
flux-transmission windows together with long absorption troughs at $2.7<z<3.0$
have been interpreted as evidence for an extended reionization process.

Indirect information on the redshift and duration of \heii\ reionization may be
also obtained from the thermal state of the IGM
\citep[e.g.][]{Miralda1994, Hui1997, Ricotti, Schaye, McDonald2001, Theuns, Lidz2010, Curvature}.
In fact, the photoionization of \heii\
may contribute substantially to the heating of the IGM,
especially at low densities,
thus resulting in an overall increase of the IGM temperature.
When reionization is completed,
the thermal evolution will asymptotically reach a state in which the
adiabatic cooling of the IGM is nearly balanced by photoheating
\citep[e.g.][]{Hui1997}.
Early estimates of the IGM temperature evolution from the width of the
\hi\ Ly$\alpha$ forest absorption lines
in quasar spectra suggested a rapid change at $z\sim3$ \citep[e.g.][]{Schaye}.
The analysis by \citet{Theuns} of the average \hi\ optical depth along the lines
of sight of about a thousand
quasars evidenced the presence of a narrow dip
at $z\sim 3.2$ in the evolution of the \hi\ optical depth.
This feature was interpreted as an effect of the increased
IGM temperature due to \heii\ reionization.
However, several subsequent studies did not confirm
this result \citep[e.g.][]{McDonald2005, Kim, DallAglio2009}.
Also, later theoretical analyses
argued that such a transition in the \hi\ optical depth would have been
too rapid to be due to IGM photoheating and adiabatic cooling alone
\citep[e.g.][]{Bolton2009b, McQuinn}.
Recently, \citet{BeckerTau} have presented a measurement of the mean IGM
temperature
evolution using a new statistics based on the curvature
of the \hi\ Ly$\alpha$
forest spectra and calibrated against numerical
simulations of spatially homogeneous \heii\ reionization.
When applied to 61 high-resolution quasar spectra,
this new statistics suggests a much more extended
and gradual increase in the mean IGM temperature between $3<z<4$.
This result is consistent with
an extended \heii\ reionization process as inferred from
the analysis of the \heii\ optical depth.

Although these methods provide a consistent picture of the onset
and global duration of \heii\ reionization,
very little is known about the detailed geometry, topology and patchiness of this process.
Indeed, past investigations based on the measurement
of the IGM temperature have focused mostly on the detection of a global signal,
while studies of the \heii\ optical depth still suffer from small-sample
statistics.
How inhomogeneous was \heii\ reionization? What are the relevant physical scales
associated with this process? Is the topology of \heii\ reionization consistent
with the current understanding of the AGN population?
In this paper, we use hydrodynamical simulations post-processed with radiative-transfer calculations
in order to address these key questions and study the impact of \heii\ reionization on to
both the \hi\ and the \heii\ Ly$\alpha$ forests.
This is a demanding task due to the multiscale nature of the reionization process.
AGNs have typical comoving separations of tens of Mpc and each of them can produce an
ionized region with a similar
characteristic size. On the other hand,
single absorption features of the \hi\ Ly$\alpha$ forest are produced on much smaller scales, $\sim 100\,\tr{kpc}$.
Resolving these structures while accounting for the statistics of the quasar population therefore
requires simulations with a large dynamic range, i.e. encompassing a sizable volume which is
partitioned into small computational elements.
Given the limitations of current computational facilities and algorithms,
previous work on the subject has mostly focused either
on small \citep[e.g.][]{Bolton2009b,Meiksin} or on large
\citep{Sokasian2002,Paschos,McQuinn} scales using various
numerical schemes to follow (or mimic) the gas dynamics and the propagation of the ionization fronts.
While high-resolution simulations within volumes with a linear size of $\sim 30\,\tr{Mpc}$
\citep[e.g.][]{Meiksin} have the advantage
of resolving the physical scales associated with the \hi\ Ly$\alpha$ forest,
they are not able to fully capture the patchiness of the \heii\ reionization process
and to study the effects of the inhomogeneous radiation field
on the \heii\ Ly$\alpha$ forest.
On the other hand, larger computational boxes
that extend for several hundreds of Mpc \citep[e.g.][]{McQuinn}
provide realistic \heii\ Ly$\alpha$ forest spectra but
necessarily lack the spatial resolution for a detailed study of the IGM heating on the \hi\ Ly$\alpha$
forest, even more so if the distribution of the baryons is `painted' a posteriori on to the dark matter
one.
We take the unprecedented step of using adaptive mesh refinement (AMR) techniques for both the
hydrodynamic and radiative-transfer calculations, which allow us to delve into the multiscale nature
of \heii\ reionization.
Specifically, to overcome the above-mentioned limitations,
we combine AMR simulations with different characteristics: large, statistically representative volumes
are analysed together with smaller, high-resolution regions which have been extracted from
larger boxes with the `zoom-in' technique.
In all cases, the hydrodynamic output is post-processed with the radiative-transfer code
{\sevensize RADAMESH} \citep{Radamesh}
which has been specifically developed to be computationally efficient on AMR
structures.
This approach
gives us the necessary dynamic range to study the imprint of \heii\ reionization
on both the \heii\ and \hi\ Ly$\alpha$ forest and to explore different statistical
tools based on the joint analysis of these two observables.

The paper is organized as follows.
We describe our simulations in Section 2 and present some basic results in Section 3 where we focus
on the \heii\ reionization history and the temperature--density relation. Simulated
absorption spectra are also presented there.
A clear prediction of our model is the presence
of an evident bimodality in the IGM temperature at fixed density during the early phases of inhomogeneous
\heii\ reionization.
In Section 4, we discuss the global \heii\ reionization signal
in terms of the \hi\ and \heii\ optical depths
together with the statistics of dark gaps and flux-transmission windows in
the \heii\ Ly$\alpha$ forest.
On the other hand, in Section 5,
we focus on smaller scales and explore the patchiness of \heii\ reionization.
In particular, we discuss proxies for the IGM temperature and present
different statistical methods
to distinguish `cold' (\heii) from `hot' (\heiii) regions in the \hi\ Ly$\alpha$ forest
during \heii\ reionization.
This study suggests that it should be possible to detect the predicted temperature bimodality with
high-resolution quasar spectra.
We summarize our results and conclude in Section 6.

\section[]{NUMERICAL METHODS}
\label{Section:simulations}

\subsection{Hydrodynamical simulations}

We perform three hydrodynamical simulations using an upgraded version of
the publicly available AMR code {\sevensize RAMSES} \citep{Teyssier}.
In all cases, we consider a flat $\Lambda$ cold dark matter cosmology
in accordance with the \ti{Wilkinson Microwave Anisotropy Probe}
results by \citet{WMAP7}.
Specifically, we adopt a
matter density parameter $\Omega_\tr{m}=0.2726$, a baryon density parameter
$\Omega_\tr{b}=0.0456$ and a present-day value of the Hubble constant
$H_0=100\,h~\tr{km}\,\tr{s}^{-1}\,\tr{Mpc}^{-1}$ with $h=0.704$.
Primordial density perturbations form a Gaussian random field
with a scale invariant spectrum
characterized by the spectral index $n=0.963$ and a linear rms value
within $8\,\tr{\mh}$ spheres of $\sigma_8=0.809$.
We assume that the gas has an ideal equation of state with adiabatic
index $\gamma=5/3$ and a chemical composition produced by big bang
nucleosynthesis with a helium mass fraction of $0.24$.
We model the ionizing radiation emitted by stars in galaxies in terms of
a time-varying but spatially uniform background for which we
adopt the recent results by \citet{HM2012}.
No contribution from AGNs is considered until
redshift $z_{\rm AGN}$. We generally adopt $z_{\rm AGN}=4$ but
we also explore an early \he\ reionization scenario in which $z_{\rm AGN}=5$.
In this work, the latter model will be only used to perform simple tests on the
reliability of our main simulations. Its detailed results will be presented in a separate paper.

We evolve the initial conditions generated with the {\sevensize GRAFIC} package
\citep{Bertschinger} from redshift $120$ to redshift $4$ accounting for photoheating
and radiative cooling. We neglect small-scale phenomena like
star formation, supernova feedback and metal enrichment, which have little
impact on the \hi\ and \heii\ Ly$\alpha$ absorption lines due to intergalactic
gas.

To account for a large statistically representative volume,
we consider a periodic cubic box of $100\,\tr{\mh}$ (hereafter dubbed the `L' box)
originally discretized into a regular Cartesian mesh with $256^3$ elements
(corresponding to a dark matter particle mass of $m_{\rm DM}\simeq 3.76 \times 10^9\,\tr{\sunh}$).
We adopt a quasi-Lagrangian AMR strategy based on local density: cells are
tagged for refinement whenever they contain more than 8 dark matter particles
or their baryonic mass exceeds a pre-fixed value.
As a result, 6 further levels of refinement on top of the base grid
are added at $z=4$ reaching an effective resolution of $16\tn 384^3$ in the
densest regions (corresponding to a cell size of $6\,\tr{\kh}$).
In what follows, we will use this box
to produce mock absorption-line spectra, which are
considerably extended in wavelength but have a relatively poor
spectral resolution.

\begin{table}
\centering
\begin{tabular}{c c r c l}
\hline\hline
Species & Bins &\multicolumn{3}{c}{Energy (ryd)} \\ [0.5ex]
\hline
\hi   & 10 & $1$    &$-$& $1.81$\\
\hei  & 10 & $1.81$ &$-$& $4$\\
\heii & 30 & $4$    &$-$& $40$\\[1ex]
\hline
\end{tabular}
\caption{Number of spectral bins and corresponding energy ranges used to sample
the spectra of AGNs.}
\label{table:energy_int}
\end{table}

\begin{table}
\centering
\begin{tabular}{c c c}
\hline\hline
Authors & Redshift & AGN lifetime\\ [0.5ex]
\hline
\citet{Haiman}         & $2\lesssim z \lesssim 4$ & $1-100$  \Myr \\
\citet{Martini}        & $z\simeq 2$              & $10-100$ \Myr \\
\citet{Jakobsen}       & $z\simeq 3$              & $>10$    \Myr \\
\citet{Porciani}       & $z\simeq 2$              & $\sim100$    \Myr \\
\citet{Fluorescence07} & $z\simeq 3$              & $\gtrsim60$  \Myr \\
\citet{ShenLifetime}   & $2.9\leq z \leq 3.5$     & $4-50$   \Myr \\
\citet{BoltonLifetime} & $z\simeq 6$              & $>3$     \Myr \\
[1ex]
\hline
\end{tabular}
\caption{Observational estimates of quasar lifetime.}
\label{table:AGNlifetime}
\end{table}

\begin{table*}
\centering
\begin{tabular}{l c c c c c l}
\hline\hline
Simulation & $L_\tr{box}$& $\Delta L_\tr{min}$ $^a$ & $\Delta L_\tr{max}$ $^b$ & AGN model & $z_\tr{AGN}$ & Comments\\
&$\tr{(\mh)}$ &$\tr{(\kh)}$ & $\tr{(\kh)}$ & & & \\[0.5ex]
\hline
L1  & $100$  & $390$ & $6$  & PLE & $4$ & Reference simulation\\
L1b & $100$  & $390$ & $6$  & PLE & $4$ & Same as L1 but with a different reionization history\\
L1E & $100$  & $390$ & $6$  & PLE & $5$ & Early ionization scenario\\
L2  & $100$  & $390$ & $6$  & PDE & $4$ & Different AGN model\\
\hline
S1H & $14~^c$& $109$ & $7$ & PLE & $4$ & High-resolution region of a $56\,\tr{\mh}$ box (S1)\\
S2H & $14~^c$& $109$ & $7$ & PLE & $4$ & High-resolution region of a second $56\,\tr{\mh}$ box (S2)\\
S2Hb& $14~^c$& $109$ & $7$ & PLE & $4$ & Same as S2H but with a different reionization history\\
\hline
LHO  & $100$ & $390$ & $6$ & Hom & $4$ & Homogeneous UV background \\
\hline
\end{tabular}
\\
\begin{flushleft}
\textit{Notes:} \\
$^a$Size of a resolution element associated with the base grid.\\
$^b$Size of a resolution element associated with the highest level of refinement.\\
$^c$Note that we neglect $1\,\tr{\mh}$ from each side of these volumes, so that the effective length is $12\,\tr{\mh}$.
\end{flushleft}

\caption{Summary of the simulations used in this work.}
\label{table:sims}
\end{table*}

In order to better resolve the low-density regions that generate the Ly$\alpha$
forest, we also perform two independent simulations with a box size of
$56\,\tr{\mh}$ adopting a zoom-in technique (dubbed the S1 and S2 boxes).
We consider three nested levels in the initial conditions with increasing
spatial resolution. The outer region has an effective resolution of
$128^3$ cells (within the entire box) while the innermost $14\,\tr{\mh}$
reach $512^3$ (corresponding to a dark matter particle mass of
$m_{\rm DM}\simeq 1.29 \times 10^6\,\tr{\sunh}$).
The AMR technique adds four additional levels in the high-resolution region
where the mesh size ranges between $109\,\tr{\kh}$ at low
densities and $7\,\tr{\kh}$ at the densest spots.
The outer layers of the high-resolution region are contaminated with
dark matter particles coming from the mid-resolution buffer zone.
For this reason, we exclude $1\,\tr{\mh}$ from each edge of the high-resolution
mesh, thus reducing its effective size to $(12\,\tr{\mh})^3$ (hereafter we will
refer to these regions as the S1H and S2H boxes).
Note that both these volumes are slightly underdense as expected for most randomly
selected volumes of relatively small size.
At $z=4$, their mean density matches
82.6 and 84.2 per cent of the cosmic value, respectively.

In all cases, we resolve the Jeans length of the gas with several computational
mesh points.

\subsection{Numerical radiative transfer}
\label{Section:Radamesh}

We model the soft UV radiation emitted by hot
stars in terms of a uniform background computed as in \citet{HM2012}.
At redshifts $z<z_{\rm AGN}$, harder radiation from AGNs
is also considered.
For the latter, we first distribute a set of point sources
throughout the entire simulation domains (see \S \ref{Section:AGN_model}
for a detailed description)
and then propagate the photons they emit by solving the radiative-transfer
equation until He reionization is completed.
As current computing facilities cannot handle the complexity of
radiation transport fully coupled to gas hydrodynamics in cosmological volumes,
photons from the discrete sources are propagated in post-processing throughout
a pre-computed snapshot of the hydro simulations (taken at $z=4$).
In Appendix \ref{hydro}, we show that
the systematic effects due to neglecting the hydrodynamical response of the
gas to photoionization heating are generally small.
This justifies our approach.

We follow the propagation of ionizing radiation using the three-dimensional
radiative-transfer code {\sevensize RADAMESH}, specifically developed to be
computationally efficient with AMR simulations and presented in \citet{Radamesh}.
The code is based on a photon-conserving
ray-tracing scheme which is adaptive in space and time
and limits the propagation of the ionization fronts to the speed of light.
The number density of six species (\hi, \hii, \hei, \heii, \heiii\ and $e^-$)
and the gas temperature are computed with a time-dependent, non-equilibrium
chemistry solver.

We sample the spectra of discrete
radiation sources using $50$ logarithmically spaced frequency bins in the
energy interval $1-40\,\tr{ryd}$, distributed as in Table
\ref{table:energy_int}
[see Appendix \ref{convergence_test} and \citet{Radamesh} for convergence
tests]. Following \citet{McQuinn},
we do not consider secondary ionizations due to X-ray photons
as nearly all the excess energy of the ionizing photons
goes into heating the gas \citep{Shull-vanSteenberg-1985}.

In order to account for the large mean free path of ionizing photons,
we propagate radiation using periodic boundary conditions.
When a photon exits the computational box for the second time, it
is excluded from subsequent calculations and added to a
homogeneous background.

The accuracy of the radiative-transfer scheme, the implicit chemistry solver, and the calculation of the gas
temperature in {\sevensize RADAMESH} depend on the time step size.
Short time steps are required to resolve the ionization history in those cells that lie in the vicinity of AGNs,
where the flux of ionizing radiation is extremely high. This could result in prohibitive execution times.
We find that the optimal trade-off between computational speed and accuracy
is achieved by using time steps
ranging between $1$ and $5\,\tr{\Myr}$. The exact value is chosen according to
the fastest evolving cell in the simulation.

The radiative transfer is performed in the L box and in the
entire $(56\,\tr{\mh})^3$ volume of the S boxes
so that we can account for the influence of distant sources on to the
high-resolution region.
A statistically representative sample of
low-resolution mock absorption spectra is extracted from the L box.
On the other hand, we use the S boxes
to achieve a spectral resolution (corresponding to $\sim 12\,\tr{\kms}$
at $z=3.5$) which is sufficient to identify typical lines of the
\hi\ Ly$\alpha$ forest \citep[e.g.][]{Curvature}.

\subsection{Sources of hard UV radiation}
\label{Section:AGN_model}
We use the {\sevensize HOP} halo finder \citep{Eisenstein} with standard parameters to identify
dark matter haloes within our simulation domains at $z=4$.
We then assume that each halo can potentially host an AGN at its centre
(the maximum halo mass in the L volume is
$1.06 \times 10^{13}\,\tr{\sunh}$ and AGNs are rare,
so it is unlikely that multiple occupancy is relevant for our work).
AGNs are randomly switched-on with a probability that
is independent of the properties of the host haloes and is related only to the
mean fraction of ionizing sources and to their
lifetime. The latter is assumed to be equal to $45$ \Myr, consistent with
the recent observational estimates summarized in Table \ref{table:AGNlifetime}.
This procedure allows us to fill our computational volumes with a population
of AGNs that realistically trace the large-scale structure formed by
gravitational instability.

We also make sure that our mock AGN population at $z=4$ reproduces the quasar
luminosity function measured by \citet{GlikmanB}.
Following \citet{Silk} and \citet{Wyithe},
we assume that the mean AB magnitude of an AGN in the $1450~\tr{\AA}$ band
depends on the mass of its host halo (as more massive haloes will
harbour more massive black holes on average):
\be \label{mass2mag}
\langle M_{1450}|M_{\rm halo} \rangle =\epsilon -2.5\,\frac{4}{3}\,\log \frac{M_{\rm halo}}{h^{-1} M_\tr{\sun}},
\ee
where $\epsilon$ is a constant.
We also assume that $M_{1450}$ follows a Gaussian distribution with rms
value $\sigma_M$.
The parameters $\epsilon$ and $\sigma_M$ are determined by imposing
that the halo mass function in the simulations is mapped
on to the quasar luminosity function by \citet{GlikmanB}.
The preferred values lie along a one-dimensional sequence: changing $\epsilon$
and $\sigma_M$ along the sequence slightly modifies the range of luminosities 
associated with our haloes.
We adopt $\epsilon=18.5$ and $\sigma_M=1.2$.
This implies that AGNs with magnitude $M_{1450} < -25$ are located in haloes
with average mass of $4.37 \times 10^{12}\,\tr{\sunh}$, in accordance with recent
results from clustering measurements \citep[e.g.][]{Porciani} and semi-analytic models
of galaxy formation \citep[e.g.][]{Fanidakis}.

\begin{figure}
\centering
\includegraphics[width=0.5\textwidth]{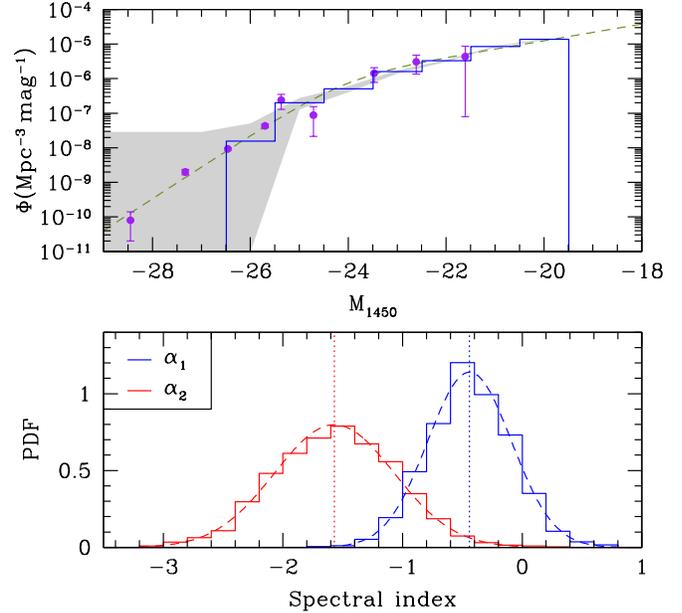}
\caption{Top: AGN luminosity function in the L box at $z=4$ (blue) compared with data points (purple) from \citet{GlikmanB}.
The shaded region shows the statistical uncertainty in the number counts for our choice of the $\epsilon$ and $\sigma_M$ parameters.
The double power-law best-fitting model of \citet{GlikmanB} is also reported for comparison (green dashed line).
Bottom: distribution of the spectral indices $\alpha_1$ (blue) and $\alpha_2$ (red) for the sources in the L box.
The dashed lines represent Gaussian distributions with mean $-0.44$ and rms value of $0.35$ (blue) and with mean $-1.57$ and rms value of $0.50$ (red).
The average of the distributions is indicated with a vertical dotted line.}
\label{lum_func}
\end{figure}

Because of the finite mass resolution in the simulations, the halo mass
function is underestimated at the low-mass end. Comparing with previous
studies \citep[e.g.][]{Jenkins}, we find that the minimum halo mass at
which the mass function in the L box is robustly determined equals
$2.5 \times 10^{11}\,\tr{\sunh}$.
This conservative limit approximately corresponds to $66$ dark matter particles.
We therefore decide to consider only AGNs with $M_{1450}\leq-19.5$ obtaining $80$ sources in the L box  at $z=4$ (out of $1805$
dark matter haloes that could potentially host AGNs fulfilling the magnitude threshold)
and $10$ in the S simulations.
Assuming that the best-fitting model to the luminosity function given in \citet{GlikmanB} holds also for (unobserved) faint AGNs, our sample would then
account for $83$ per cent of the total emissivity of AGNs at $z=4$. 
Note, however, that it includes all the observed sources (see 
top panel of Fig. \ref{lum_func}).

We approximate the spectral energy distribution of the AGNs with
a broken power law
\be
L_\nu \propto \left\{
\begin{array}{l l}
\nu ^{\alpha_1} & \quad \lambda \geq 1300 ~\tr{\AA}\\
\nu ^{\alpha_2} & \quad \lambda < 1300 ~\tr{\AA},
\end{array}
\right.
\ee
where the spectral indices $\alpha_1$ and $\alpha_2$ are randomly drawn from
two independent Gaussian distributions with mean and rms value of
$-0.44\pm 0.35$ \citep{Vanden} and $-1.57\pm 0.50$ \citep{Telfer}, respectively (see bottom panel of Fig. \ref{lum_func}).

To account for the redshift evolution of the sources, we compute
the relative change of the AGN monochromatic emissivity, $I(z)$ (number of photons
with an energy of
$13.6$ eV released per unit time per unit comoving volume),
with respect to redshift 4 using equation 37 in \citet{HM2012}:
$R(z)=I(z)/I(z=4)$ (see Fig. \ref{PLEvsPDE_evol}).
This only determines a global property of the AGN population while
we need information for each single source to perform the radiative transfer.
A further complication is that
our halo population is identified in the snapshot at $z=4$,
where we also assign AGN magnitudes to the haloes.
For these reasons, we consider two approximated schemes.
In the first one, that we dub Pure Luminosity Evolution or PLE,
we simply modify the luminosity of each newly active source that
turns on at redshift $z$ by a factor $R(z)$
(with respect to what would be assigned to the same halo at $z=4$)
without altering the mean
number of AGNs present in the simulated volume at a given time.
As an alternative, we modify the average number of active sources in the box
by a factor $R(z)$ keeping the potential luminosity of each halo fixed as
determined at $z=4$. We call this scheme Pure Density Evolution or PDE.

Finally, we also perform a simulation of homogeneous reionization in the L box
using the photoionization and photoheating rates reported in \citet{HM2012}.

The main characteristics of our simulations are summarized in Table
\ref{table:sims}.

\section[]{RESULTS}
\subsection{Initial conditions at $z=4$}

\begin{figure}
\centering
\includegraphics[width=0.5\textwidth]{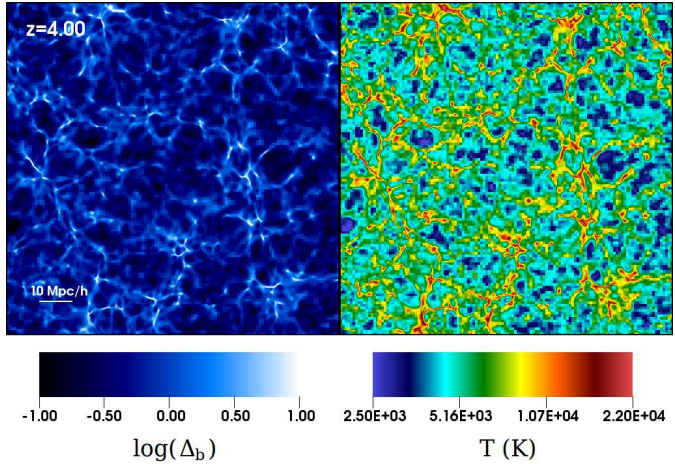}
\caption{Overdensity (left) and temperature (right) distribution in a
slice of the L1 simulation at $z=4$.}
\label{Dens_and_T}
\end{figure}

Before presenting our main results on helium reionization, we briefly
introduce the output of the hydro simulations at $z=4$ that form the
starting point of our radiative-transfer calculations.
Due to the presence of the galactic component of the UV background,
hydrogen is nearly completely ionized
with an \hi\ filling factor (the volume-weighted average of the local
\hi\ fraction) of $\sim 2 \times 10^{-5}$
while helium is
mostly in the form of \heii. Less than $0.1$ per cent of helium atoms are
ionized twice due to collisions while the fraction of neutral atoms is
smaller than $\sim 4 \times 10^{-5}$.
In Fig. \ref{Dens_and_T}, we show
the density distribution of the baryonic material
(plotted in terms of the overdensity
$\Delta_{\rm b}(\boldsymbol{x})=\rho_{\rm b}(\boldsymbol{x})/\langle\rho_{\rm b}\rangle$)
and the temperature in a slice of the L1 simulation.

\begin{figure}
\centering
\includegraphics[width=0.5\textwidth]{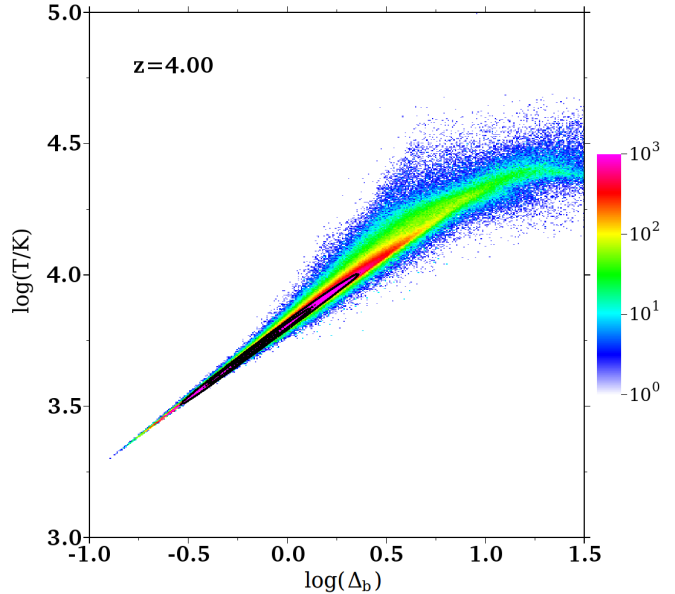}
\caption{Temperature--density relation for low-density IGM in the L1 simulation at $z=4$
when the volume filling factor of \heiii\ is less than $0.1$ per cent.
The colour scale
indicates the gas mass per pixel (in arbitrary units normalized to 1000 at maximum).
Overplotted contours correspond to 10, 30, 50, 70 and 90 per cent of the maximum value.
Note that the different levels are hardly distinguishable as
most of the gas follows a very tight temperature--density relation.
}
\label{phase_d}
\end{figure}

At low densities, the IGM shows a tight temperature--density
relation (see Fig. \ref{phase_d})
determined primarily by the balance between photoheating, adiabatic
cooling/heating and Compton cooling
\citep[e.g.][]{Miralda1994, Hui1997}.
As a result of these effects, we can write
an `effective equation of state' of the form
\be
T=T_0\,\Delta_\tr{b}^{\gamma -1},
\label{eeos}
\ee
where the gas temperature at mean density, $T_0 \simeq 6300\,\tr{K}$, and the polytropic index,
$\gamma\simeq 1.56$, are determined by the cosmological model and
the ionization history of the gas. The scatter around this relation
is very small, i.e. the rms temperature variation at $\Delta_\tr{b}=1$ is $\sigma_0\simeq 140\,\tr{K}$.

\begin{figure*}
\centering
\includegraphics[width=1.\textwidth]{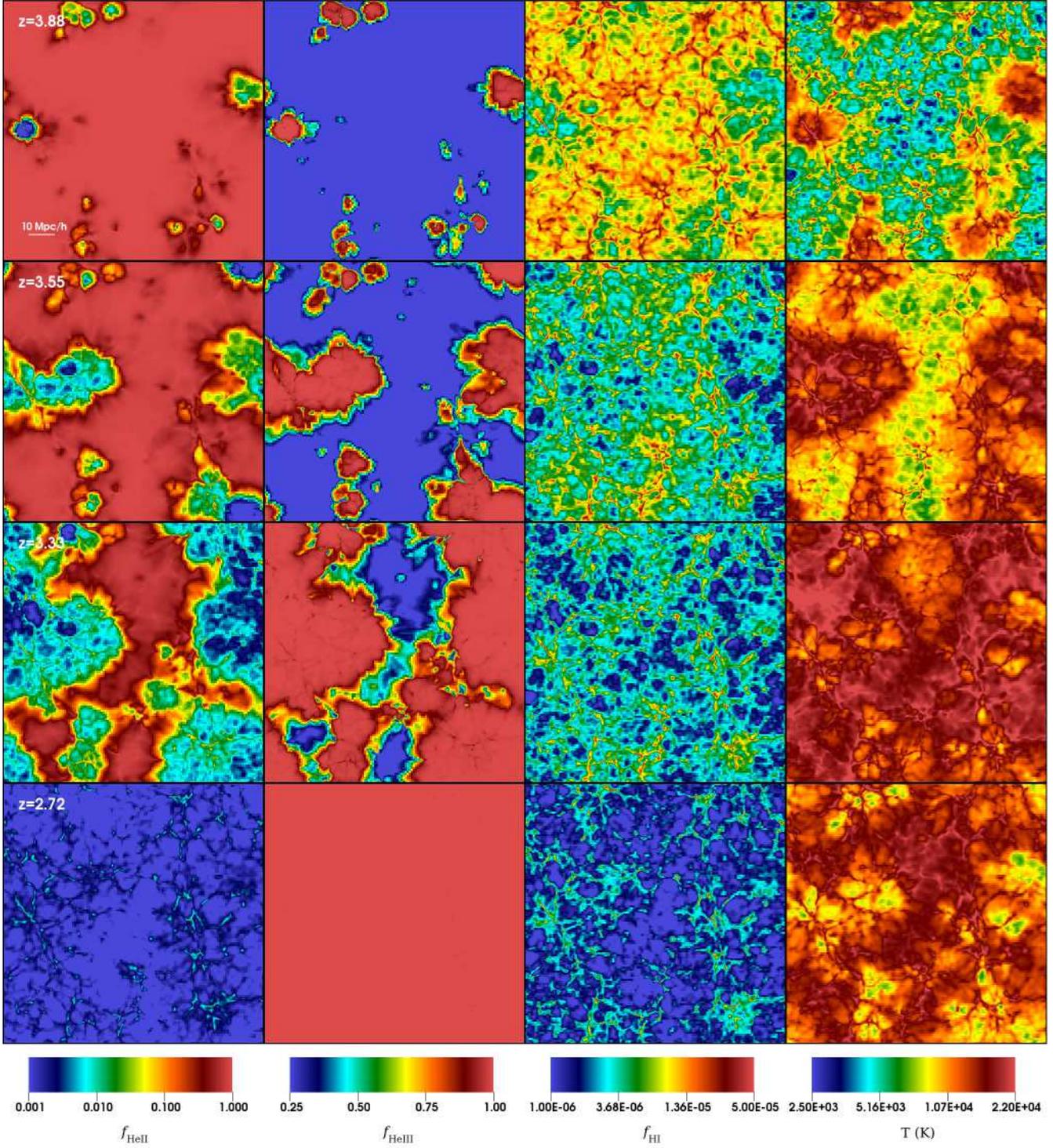}
\caption{Slices through the L1 simulation. Each column refers to a different
physical quantity: the fraction of singly ionized helium ($f_{\tr{\heii}}$, left),
the fraction of \heiii\  ($f_{\tr{\heiii}}$, centre left),
the fraction of neutral hydrogen ($f_{\tr{\hi}}$, centre right) and the IGM temperature (right).
Each row, instead, corresponds to a different stage of \heii\ reionization at redshift
3.88 (top), 3.55 (top-centre), 3.33 (bottom-centre) and 2.72 (bottom) corresponding to
an \heiii\ filling factor, $f_{\tr{\heiii}}$, of $0.10,~0.51,~0.80$ and $~0.99$, respectively.}
\label{PLEslices}
\end{figure*}

\subsection{Reionization history}
\label{Section:ReionHistory}

A cartoon sketching the propagation of ionizing radiation from AGNs
through the L1 simulation is shown in Fig. \ref{PLEslices}.
This refers to the same slice presented in Fig. \ref{Dens_and_T}.
Each row corresponds to a given epoch
($z=3.88, 3.55, 3.33$ and $2.72$ from top to bottom)
and is characterized by
a specific value of the total \heiii\ filling factor
($0.10, 0.51, 0.80$ and $0.99$, respectively).
On the other hand, each column refers to a different property of the
IGM (from left to right: the \heii\ fraction,
the \heiii\ fraction, the \hi\ fraction and the gas temperature).

When an AGN turns on in a medium devoid of \heiii, it carves during its lifetime
an \heiii\ region in the IGM
with a characteristic size of several \mh\ (depending on the
luminosity of the ionizing source) before switching off.
The ionized bubbles have irregular shapes mainly determined by the
underlying distribution of matter. Ionization fronts propagate faster at
low densities,
where the time-scale for recombination is larger, and thus rapidly sweep
the volume-filling underdense regions.
At first, \heiii\ bubbles are well separated by extended
\heii\ regions. When their central source switches off, the internal \heiii\
fraction decreases due to recombinations
until a new AGN becomes active in the surroundings.
After different generations of AGNs have shined, bubbles overlap
and the \heiii\ phase percolates.
At this point, the topology
of the different ionized states of helium has reversed:
small pockets of \heii\ -- which are shielded from ionizing radiation
by high-density clumps -- emerge from an `ocean' of \heiii.

\begin{table*}
\centering
\begin{tabular}{c c c c c c c}
\hline\hline
  & \multicolumn{3}{c}{$f_\tr{\heiii}\leq 0.1$} & \multicolumn{3}{c}{$f_\tr{\heiii}\geq 0.9$} \\ [0.5ex]
$z$ & $\gamma$ & $\log (T_0/\tr{K})$ & Vol & $\gamma$ & $\log (T_0/\tr{K})$ & Vol \\
\hline
$3.77$ & $1.4412\pm 0.0001$ & $3.8506\pm 0.0002$ & $0.65$           & $1.2625\pm 0.0001$ & $4.1723\pm 0.0007$ & $0.08$ \\
$3.59$ & $1.3568\pm 0.0001$ & $3.9006\pm 0.0002$ & $0.28$           & $1.2408\pm 0.0001$ & $4.1847\pm 0.0004$ & $0.19$ \\
$3.48$ & $1.3148\pm 0.0001$ & $3.9372\pm 0.0004$ & $0.06$           & $1.2104\pm 0.0001$ & $4.2058\pm 0.0003$ & $0.36$ \\
$3.33$ & $1.2834\pm 0.0003$ & $3.9808\pm 0.0013$ & $2\cdot 10^{-3}$ & $1.1977\pm 0.0001$ & $4.2145\pm 0.0003$ & $0.61$ \\
$3.15$ & $1.42\pm 0.10$     & $3.93\pm 0.21$     & $5\cdot 10^{-5}$ & $1.1958\pm 0.0001$ & $4.2014\pm 0.0003$ & $0.78$ \\
$2.94$ & $-$                & $-$                & $0$              & $1.2048\pm 0.0001$ & $4.1906\pm 0.0003$ & $0.98$ \\
\hline
\end{tabular}
\caption{To emphasize and quantify the bimodal distribution of the gas temperature that forms during \heii\ reionization in the L1 simulation, we fit the temperature--density relation of highly ionized gas ($f_\tr{\heiii}\geq 0.9$) and lowly ionized gas ($f_\tr{\heiii}\leq 0.1$) in the density range $-1.5\leq \log(\Delta_\tr{b}) \leq 0.5$ with two separate power laws of the form given in Equation (\ref{eeos}). The volume fraction in the two ionization states is also given.}
\label{table:phase_d}
\end{table*}

\begin{figure}
\centering
\includegraphics[width=0.5\textwidth]{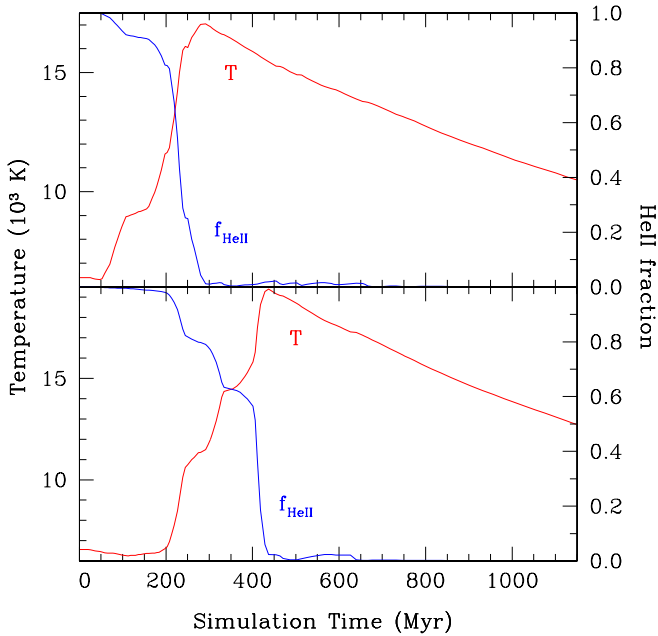}
\caption{Temperature evolution (red curve, left-hand axis) and evolution of the fraction of \heii\
(blue curve, right-hand axis) in terms of the simulation time for two different cells (top and bottom panels)
at mean density in the L1 simulation.}
\label{mean_cell}
\end{figure}

\begin{figure}
\centering
\includegraphics[width=0.5\textwidth]{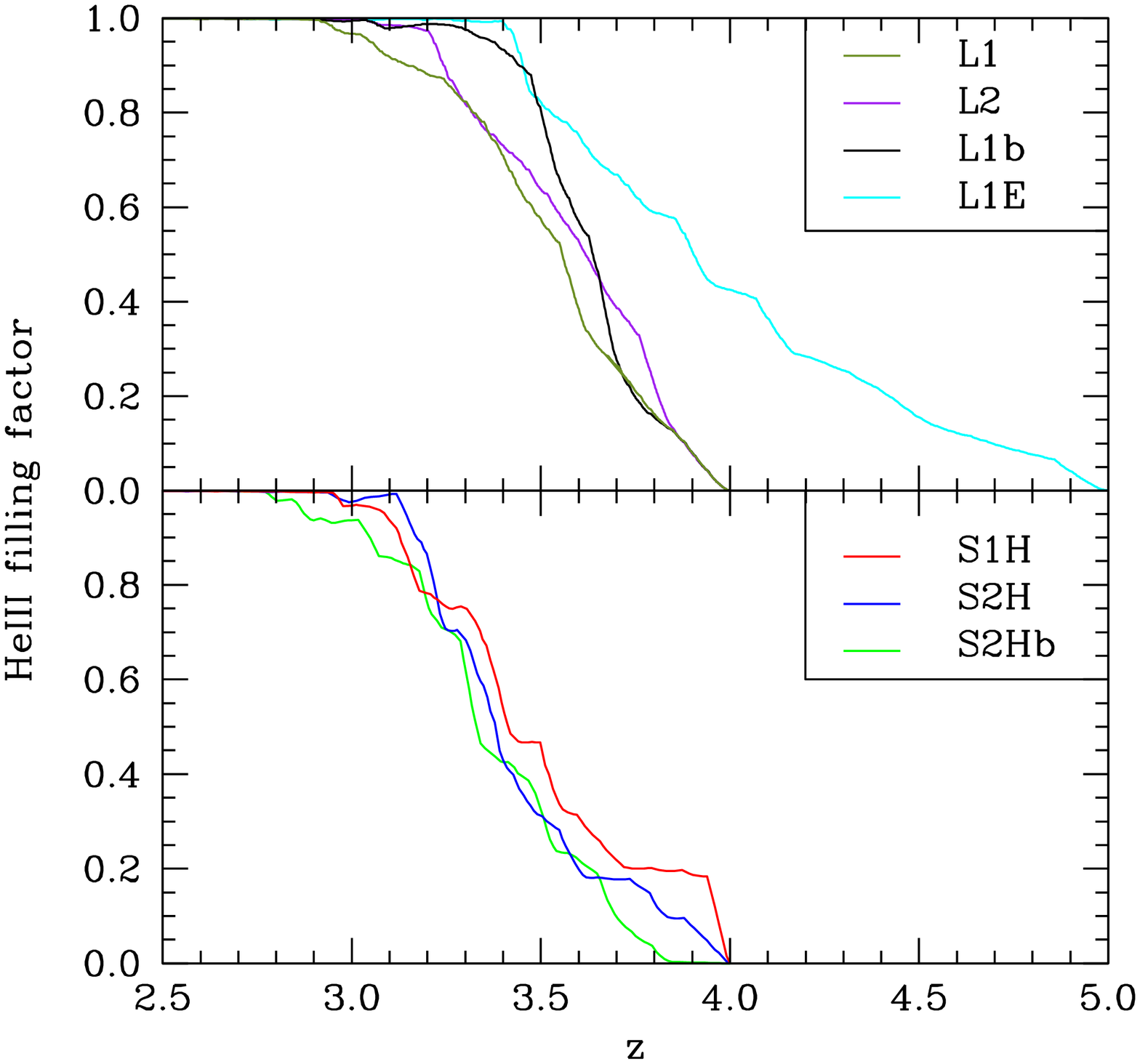}
\caption{Redshift evolution of the \heiii\ filling factor in our simulations. Top: L1 (green), L1b (black), L1E (cyan) and L2 (purple) simulations. Bottom: high-resolution regions probed by the S1H (red), S2H (blue) and S2Hb (green) volumes.}
\label{reion_trend}
\end{figure}

The passage of an ionization front is also associated with an increase in the
temperature of the medium due to the thermalization of the photoelectrons.
As an example, in Fig. \ref{mean_cell} we show the evolution of the
\heii\ fraction and of the gas temperature in
two different cells at mean density in the L1 simulation.
The top panel shows that helium is singly ionized at a temperature
of $T\sim 6400\,\tr{K}$ when
a front reaches the cell. The \heii\ fraction rapidly decreases
and the transition to a completely ionized state requires
$\sim 250\,\tr{\Myr}$. Since this time exceeds the lifetime of a single AGN,
the ionization process takes place because of the contribution
of multiple sources that switch on at different times.
A consequence of this phenomenon is evident in the figure:
at the simulation time of $\sim 100$ Myr, a nearby source switches off. Both the
temperature and the \heii\ fraction thus
reach a plateau that lasts for a few \Myr, i.e. until the ionizing photons
from other sources reach the cell
\footnote{A similar behaviour is seen for many gas elements, especially
around mean density. Substantially underdense cells are simultaneously
exposed to several sources and are quickly ionized.
On the other hand, overdense cells tend to be closer to ionizing sources
and present a variety of ionization time-scales depending on the local AGN
density.}.
Later on, ionization equilibrium is
established and the temperature reaches a maximum value
of $T\sim 17\tn000\,\tr{K}$. After the front has swept the cell,
the temperature decreases monotonically according to Hubble cooling
with a time-scale of $\sim 1$ Gyr.
The bottom panel of Fig. \ref{mean_cell} depicts a similar scenario
where the contribution from multiple sources is necessary to completely ionize
the cell.

\begin{figure*}
\centering
\includegraphics[width=1.\textwidth]{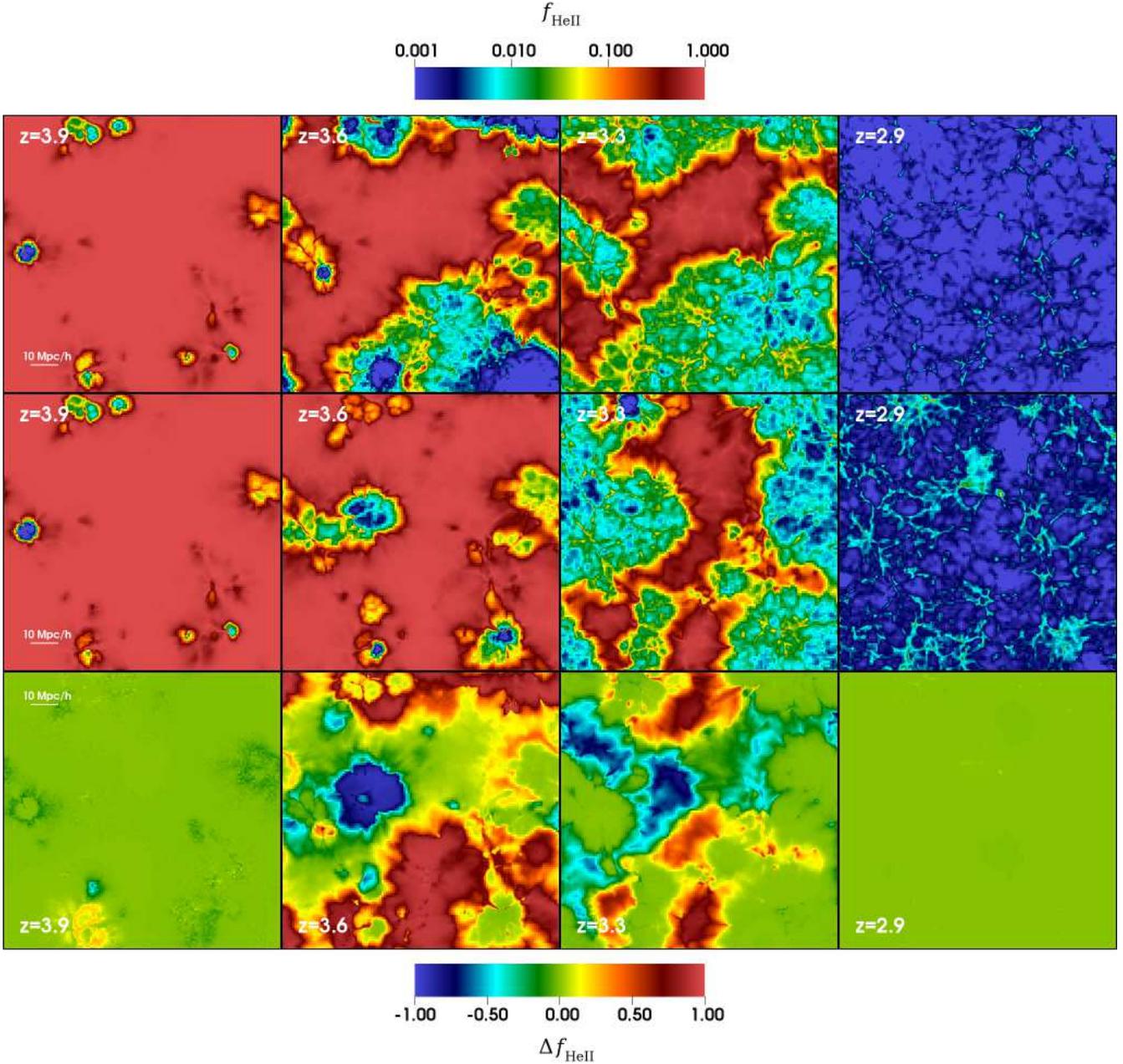}
\caption{Comparison between the source models adopted in the L1 and L2 simulations at different redshifts
(as indicated in the labels). Top: distribution of the \heii\ fraction in a slice from the L2 simulation. Centre: distribution of the \heii\ fraction in a slice of the L1 simulation as also shown in Fig. \ref{PLEslices}. Bottom: difference between the \heii\ fractions in the L1 and L2 simulations, $\Delta f_\tr{\heii} = f_\tr{\heii}^\tr{(L1)}-f_\tr{\heii}^\tr{(L2)}$.
Note that, to ease comparison, both simulations start with the same distribution of sources at $z=4$.}
\label{topology}
\end{figure*}

In Fig. \ref{reion_trend}, we show the time evolution
of the \heiii\ filling factor in the simulations. In the L series,
the filling factor approaches unity between $2.9\lesssim z \lesssim 3.1$,
in good agreement with high-quality Cosmic Origin Spectrograph observations
\citep[e.g.][]{Shull, Worseck}. Even in the L1b box, where He reionization proceeds
at a faster pace due to the presence of very luminous sources,
the fraction of \heii\ drops below $1$ per cent only at $z \simeq 3.0$,
analogously to what happens in the L1 and L2 simulations.
Overall, \heii\ ionization and the corresponding heating
of the IGM are extended processes that take place within a redshift interval
$\Delta z \gtrsim 1$
in line with the observational evidence based on the curvature of quasar
absorption lines presented in \citet{Curvature}.
This result is even more evident in the L1E simulation where the \heiii\ filling factor at $z=4$
is equal to $0.425$ but the reionization is completed only at $z< 3.4$.

Additional insight on how \heii\ reionization takes place at a local level
is provided by the S simulations.
In fact, the S1H, S2H and S2Hb boxes provide us with a few
realizations of a small portion of the Universe, characterized by
different reionization histories.
The detailed evolution of the \heiii\ filling factor depends on the actual distribution of AGNs around
and within the boxes.
For instance, at $z \sim 3.8$ roughly $20$ per cent of \he\ in the S1H volume is already fully
ionized while in the S2Hb box $f_\tr{\heiii}<0.05$.

\begin{figure*}
\centering
\includegraphics[width=0.9\textwidth]{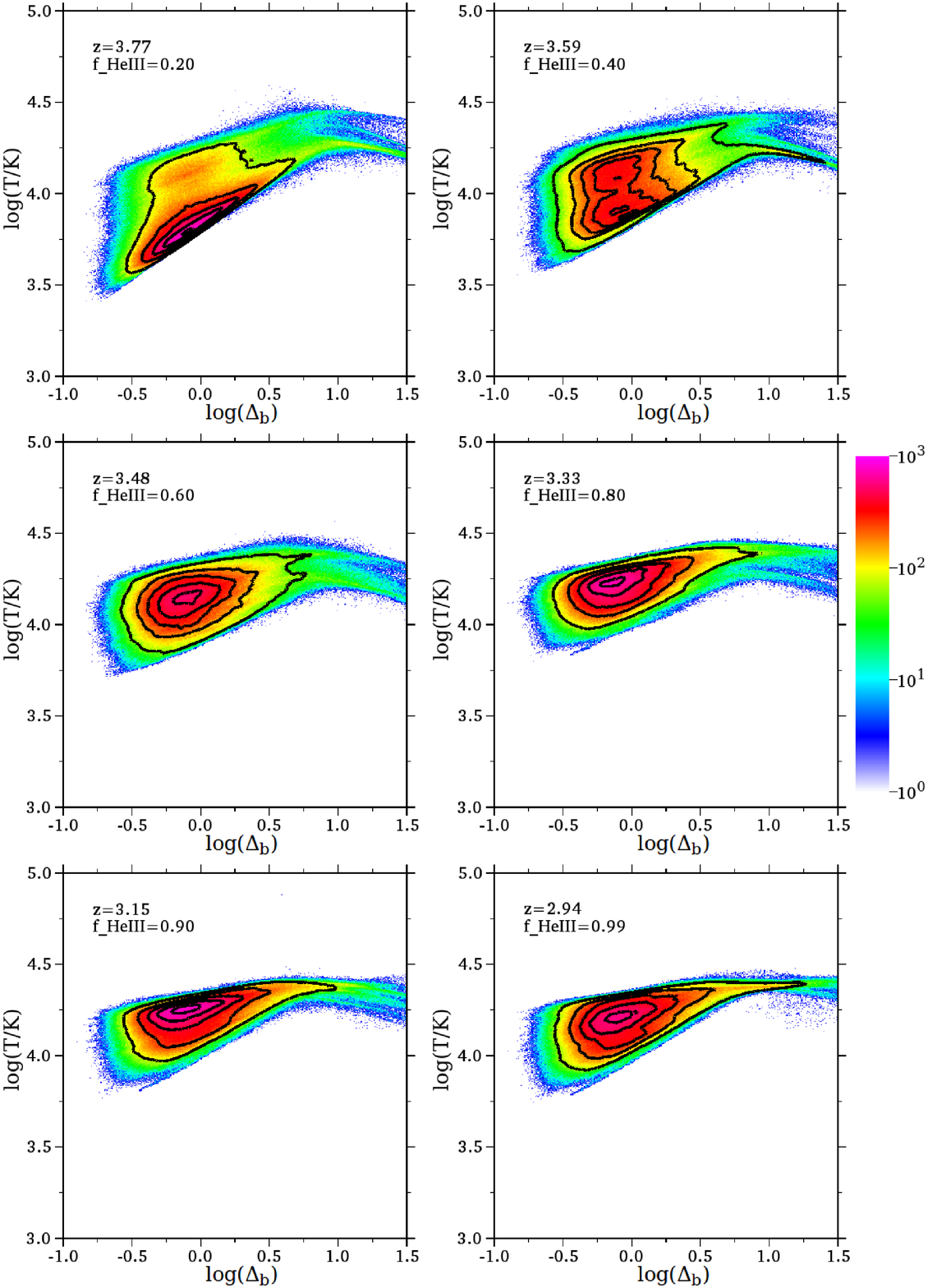}
\caption{Evolution of the temperature--density relation during \heii\ reionization in the L1 simulation.
The style of the plot is as in Fig.~\ref{phase_d}.}
\label{phase_d_PLE}
\end{figure*}

\begin{figure*}
\centering
\includegraphics[width=0.6\textwidth, angle=270]{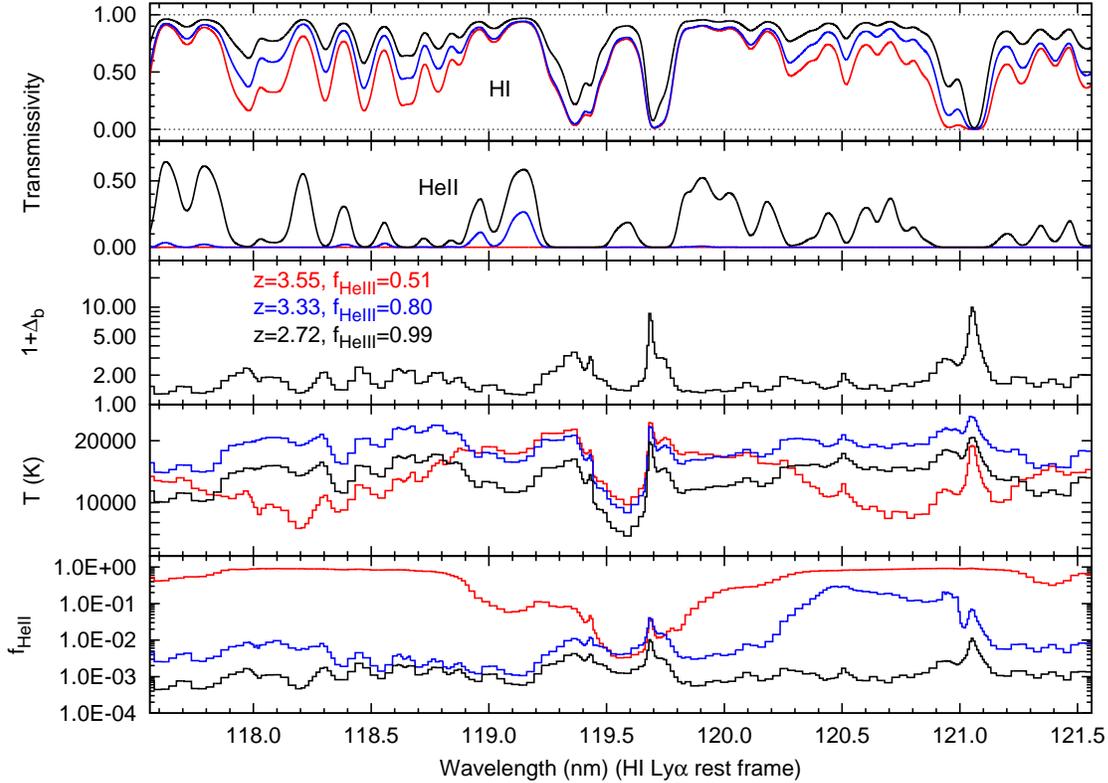}
\caption{Typical \hi\ and \heii\ Ly$\alpha$ absorption spectra along a line of sight extracted from our L1 simulation for three different redshifts: $z=3.55$ (red), $3.33$ (blue) and $2.72$ (black). From
top to bottom, we show, respectively: the \hi\ and \heii\ transmitted flux (both convolved with a Gaussian
profile with a $\tr{FWHM}$  of $88\,\tr{\kms}$), the gas overdensity, the
gas temperature and the fraction of singly ionized helium, $f_\tr{\heii}$.
The wavelength along the $x$-axis has been rescaled to the \hi\ Ly$\alpha$ transition in the rest frame
of the gas appearing at the extreme right of the plot.}
\label{spectrum1}
\end{figure*}

\begin{figure}
\centering
\includegraphics[width=0.47\textwidth]{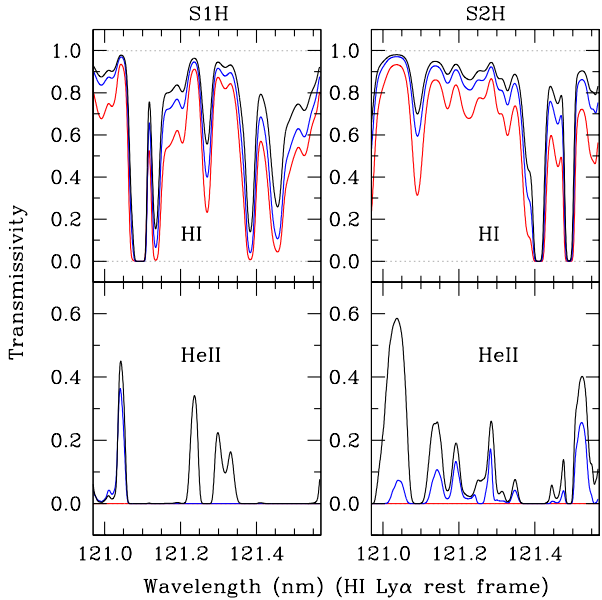}
\caption{Sample \hi\ and \heii\ Ly$\alpha$ absorption spectra along a line of sight extracted from the S1H (left) and the S2H (right) simulations at different redshifts: $z=3.5$ (red),
$3.2$ (blue) and $2.9$ (black). The synthetic spectra have been convolved with a Gaussian
profile with a $\tr{FWHM}$ of $12\,\tr{\kms}$.}
\label{spectraS}
\end{figure}

\begin{figure*}
\centering
\includegraphics[width=0.60\textwidth, angle=270]{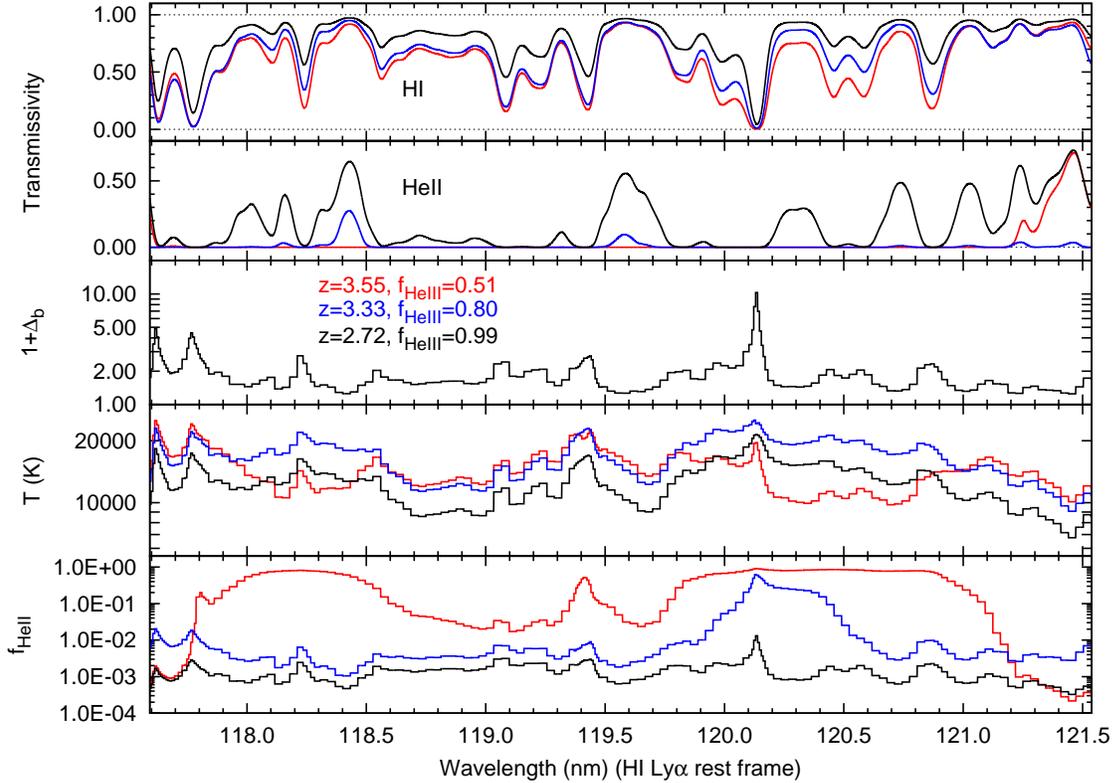}
\caption{As in Fig. \ref{spectrum1} but along a particular line of sight
which has impact parameters $\simeq 4.7$ and $\simeq 7.2\,\tr{\mh}$ with respect to a pair of closely spaced ($\sim 4.2\,\tr{\mh}$) sources active between $z=3.6$ and $3.52$ with magnitude $M_{1450}\simeq -22.4$ and $-22.9$, respectively. An intense transmission peak in the \heii\ spectrum is evident around $\lambda\simeq 121.4~\tr{nm}$ both at redshift $z=3.55$ and $2.72$ while it is not present at intermediate redshifts.}
\label{spectrum2}
\end{figure*}

\begin{figure}
\centering
\includegraphics[width=0.5\textwidth]{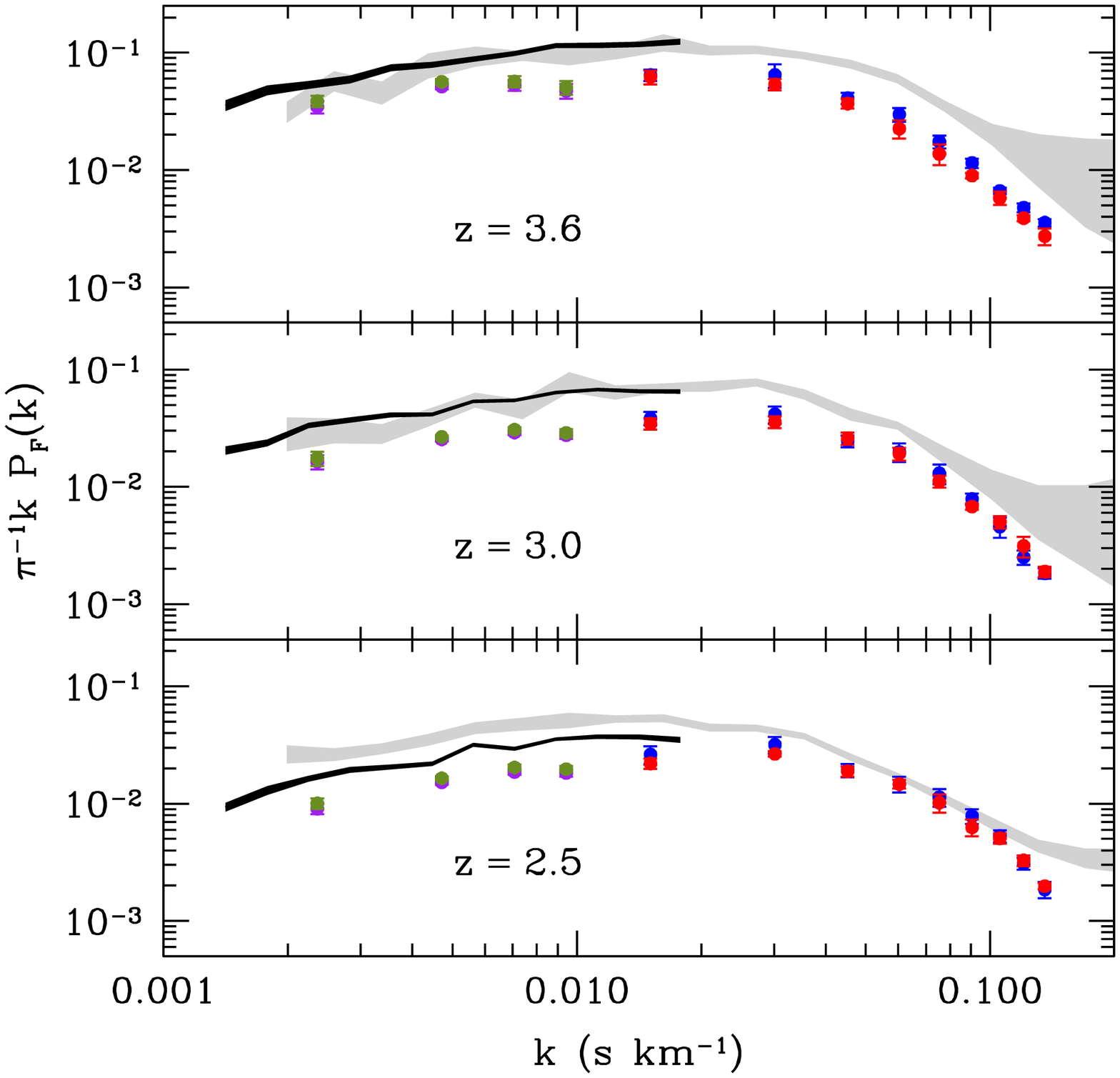}
\caption{Line-of-sight power spectrum of the \hi\ transmitted flux at different redshifts.
Points with $1\sigma$ error bars have been extracted from our simulations and different colours refer to
the various computational boxes: L1 (green), L2 (purple, hardly visible because almost perfectly overlapping with the green points), S1H (red) and S2H (blue).
The shaded regions mark the observational constraints by
\citet[][black]{McDonald2006} and \citet[][grey]{Croft}.}
\label{ps}
\end{figure}

Contrary to the global ionization history,
the detailed geometry of the reionization process (e.g. the
size of the ionized bubbles and their spatial distribution) depends on the
model which is adopted for the evolution of the ionizing sources.
This is emphasized in Fig. \ref{topology}, where we compare the \heii\
distribution in the same slice of the L1 and L2 simulations as a function of
redshift.

The PLE model, which contains fewer but more luminous sources at $z<4$, at early times
presents on average more extended ionization fronts than the PDE model, which is
characterized by smaller and more numerous ionization bubbles.
During the initial stages of the reionization process the discrepancies
are due to the properties and the location of the active sources at that
particular time step. At later times, when the bubbles
start percolating, the differences between the two models
are less evident, as shown in the third column of Fig. \ref{topology}.
Both scenarios end up producing nearly complete \heii\ ionization
at $z\sim 2.9$.

\subsection{Temperature--density relation}

Line-fitting studies of absorption features in quasar spectra
indicate that the temperature--density relation of the IGM
is characterized by a `normal' equation of state with $\gamma>1$
\citep{Bryan, Ricotti, Schaye, McDonald2001, Zaldarriaga, Rudie}.
On the contrary, using the probability density function (PDF) of the
transmitted flux, wavelet analysis, or a combination of both, other authors
find evidence for an `inverted' equation of state with $\gamma<1$
\citep{Bolton2008, Viel, Calura, Garzilli}.

In Fig. \ref{phase_d_PLE}, we show how the temperature--density relation
of the IGM is altered as \heii\ reionization progresses in the L1 box.
As discussed above,
gas elements are rapidly heated when they are invested by a photoionization
front. Therefore, while \heii\ reionization is ongoing, an ever increasing
fraction
of gas leaves the original temperature--density relation and
form a new, hotter and less tight sequence ($\sigma_0\simeq 3300\,\tr{K}$ at $z=3.5$).
As reionization proceeds, two well-defined and unequally populated $T$--$\Delta_\tr{b}$
relations are present.
We fit two different polytropic relations in the density interval $-1.5\leq \log (\Delta_\tr{b})\leq 0.5$
by separately considering gas elements with a low ($f_\tr{\heiii}\leq 0.1$) and a high ($f_\tr{\heiii}\geq 0.9$) level of ionization. The best-fitting values
for $\gamma$ and $T_0$ are reported in
Table \ref{table:phase_d} together with the volume fraction of the gas in the two ionization states
for all the redshifts appearing in Fig. \ref{phase_d_PLE}.
Eventually, when reionization is complete at $z\simeq 2.9$, a new $T$--$\Delta_\tr{b}$ relation emerges
with a value of $\gamma\simeq 1.20$ which is smaller than the initial one. At this time the gas temperature
at mean density is $T_0\sim 15\tn600\,\tr{K}$ and the rms temperature variation decreases to
$\sigma_0\simeq 2800\,\tr{K}$.
At low densities, we find no evidence for an inverted equation of state of the IGM (e.g. $\gamma<1$)
while we detect it for early times and $\Delta_{\rm b}>10$ as expected from arguments of thermal balance
\citep[e.g.][]{Meiksin1994,Miralda1994,Tittley2007,Meiksin}.

\subsection{Simulated spectra}

We produce mock Ly$\alpha$ absorption spectra based on the distribution
and velocity of \hi\ and \heii\ along different lines of sight in
our computational volumes.
We use resolution elements of $1\,\tr{\kms}$ and select lines
of sight which are parallel to one of the main axes of the computational box.
In order to mimic the instrumental response of a real spectrograph,
we convolve each spectrum with a Gaussian profile.
For the S1H, S2H and S2Hb boxes,
we use a kernel with a full width at half-maximum (FWHM) of $12\,\tr{\kms}$.
On the other hand, we adopt $88\,\tr{\kms}$
for the L1, L1b, L1E, L2 and LHO simulations which have lower spatial resolution.
In the latter case, our synthetic spectra are not able to resolve single
\hi\ lines of the Ly$\alpha$ forest but still give
information on the coarse-grained properties of the medium.

Sample spectra corresponding to the same line of sight in the L1 simulation
at different redshifts are shown in Fig. \ref{spectrum1}.
To ease the comparison, wavelengths are given for the \hi\ Ly$\alpha$
transition in the rest frame of the gas that appears at the extreme right
of the plot.
Together with the \hi\ and the \heii\ absorption spectra, the
density, temperature and \heii\ ionization fraction of the gas
are also displayed. Since we perform the radiative transfer in post-processing
on a single snapshot of the AMR simulations,
the density field evolves simply as $(1+z)^{3}$ and overdensities do not change.
On the other hand, temperatures vary with time showing
the expected behaviour: the passage of an
\heii\ ionization front rapidly heats up the medium that then
smoothly cools down.
Overall, helium second ionization makes the IGM less opaque to
both \hi\ and \heii\ Ly$\alpha$ photons.
As expected, the spectra of the two
elements present some qualitative differences.
In the redshift range, $3.55<z<2.72$, \hi\ spectra show a constant increase in
the transmitted flux but preserve the same pattern of absorption features.
On the contrary, \heii\ spectra show a much more dramatic change transiting from
complete absorption (at $z=3.55$) to the appearance of small regions of
transmitted flux (at $z=3.33$) that subsequently increase in size and
number as \heii\ reionization proceeds.
Spectra obtained in the S1H and S2H volumes present similar features
but resolve single \hi\ absorption lines, as shown in Fig. \ref{spectraS}.

In Fig. \ref{spectrum2}, we focus on a particular line of sight
in the L1 simulation which has been selected because of its vicinity to
a pair of closely spaced ($\sim 4.2\,\tr{\mh}$) luminous sources.
It has an impact parameter
of $\simeq 4.7\,\tr{\mh}$ with respect to a source with magnitude
$M_{1450}\simeq -22.4$ and $\simeq 7.2\,\tr{\mh}$ with respect to
a second AGN with magnitude $M_{1450}\simeq -22.9$
that are active between $z=3.60$ and $3.52$.
For this reason, at redshift $z=3.55$ (red line), a transmission peak in the
\heii\ spectrum is present around $\lambda\simeq 121.4~\tr{nm}$. After the
sources switch off, \heiii\ starts recombining, thus
reducing the transmissivity in the spectrum.
In the absence of additional sources of ionizing radiation
in the surroundings, the medium
recombines enough to produce an almost complete absorption.
This effect occurs at $z=3.33$ (blue line) when the
transmission peak is extremely weak. Only at $z=2.72$, when
reionization is completed, the fraction of \heii\ is again small enough to
allow some flux to be transmitted.
We will discuss extensively the implications of this kind
of features in Section \ref{ftw}.

\subsubsection{Flux power spectrum}
Line-of-sight power spectra of the \hi\ transmitted flux extracted from our simulations
are shown in Fig. \ref{ps} together with the observational data by
\citet{McDonald2006} and \citet{Croft}.
Our results are in qualitative agreement with observations and reproduce the expected shape of the power spectrum at all redshifts.
The overall amplitude of the simulated spectra slightly underestimates the observational results. This might be a consequence of the fact that
our S boxes are mildly underdense while hydrogen in the L boxes tends to be slightly more ionized than expected (see Section  \ref{optical_depths}).
Note that the L1 and L2 boxes produce nearly identical power spectra as the differences in their AGN models
do not affect the largest scales.

\section[]{GLOBAL REIONIZATION SIGNAL}
\subsection{Effective optical depth} \label{optical_depths}

\begin{figure*}
\centering
\includegraphics[width=0.49\textwidth]{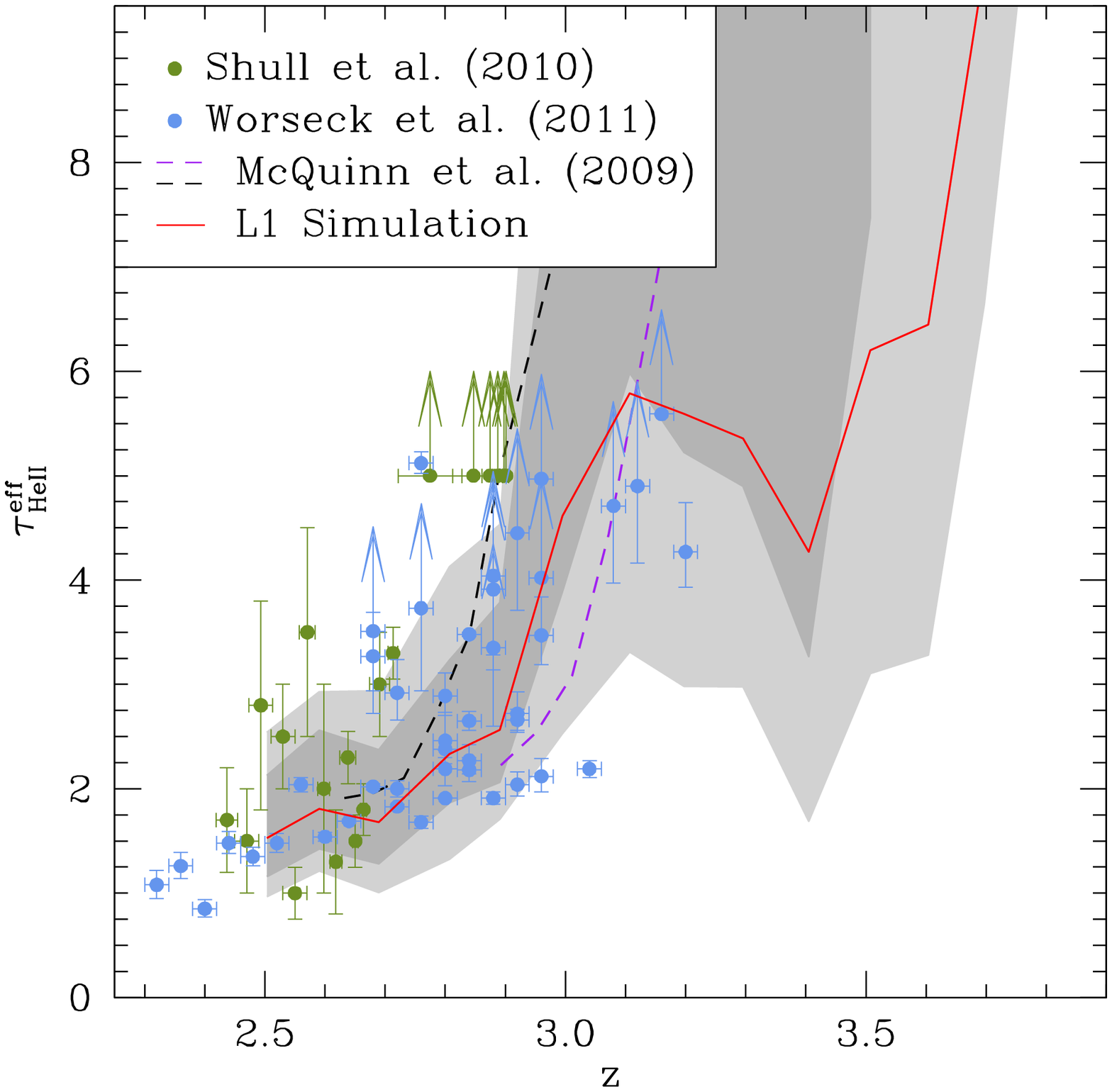}
\includegraphics[width=0.49\textwidth]{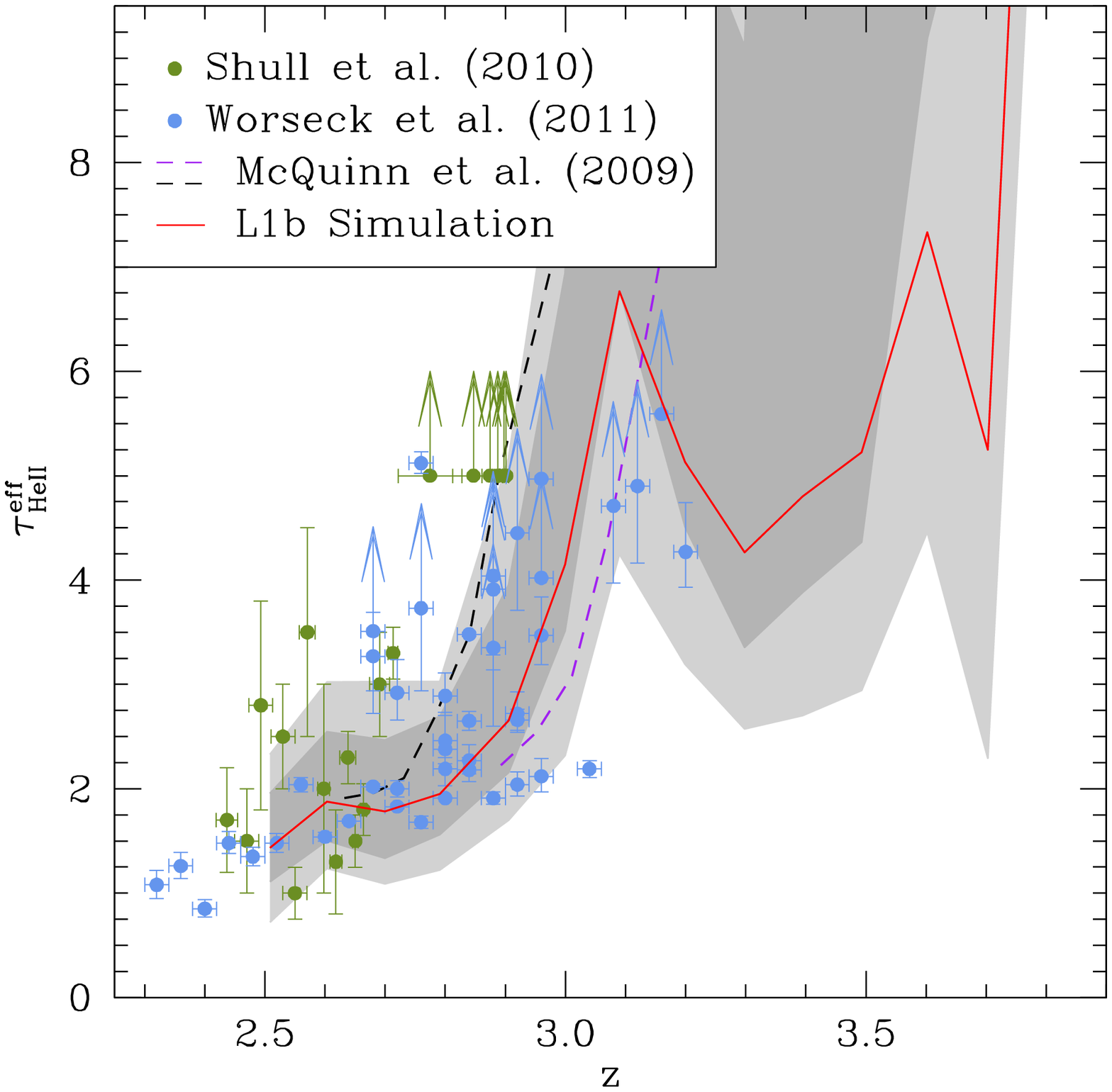}
\caption{Redshift evolution of the mean \heii\ effective optical depth (red line) evaluated in identical bins of $\Delta z=0.04$ in the L1 (left) and L1b (right) simulations. 
Shaded areas denote the regions of the $\tau_\tr{\heii}^\tr{eff}$ distribution enclosed between the $16\tr{th}$ and $84\tr{th}$ percentile (dark grey) and between the $2\tr{nd}$
and $98\tr{th}$ percentile (light grey).
Observational data by \citet{Shull} (green points, $\Delta z=0.1$) and \citet{Worseck} (blue points, $\Delta z=0.04$) are overplotted for comparison. The dashed lines indicate the mean $\tau^\tr{eff}_\tr{\heii}$ extracted from the radiative-transfer simulations 
by \citet{McQuinn} (their L3 simulation is drawn in black and their D1 simulation in purple) as computed in \citet{Worseck} with $\Delta z=0.04$.
The dip at $z\sim 3.4$ in the L1 volume is generated by a positive fluctuation in the total number of ionizing photons. Similarly, the L1b simulation presents a deficit of ionizing photons at $z\sim 3.1$ that momentarily increases the effective optical depth in the volume (see Appendix \ref{PLEvsPDE} for details).}
\label{EffTauHe_L1}
\end{figure*}

\begin{figure}
\centering
\includegraphics[width=0.5\textwidth]{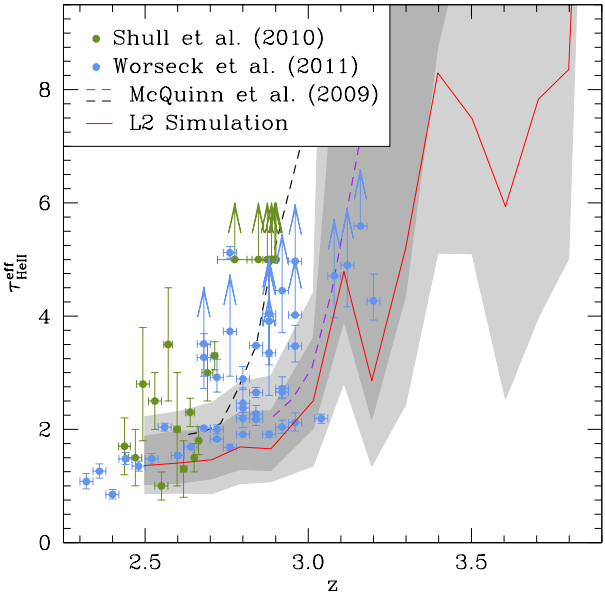}
\caption{As in Fig. \ref{EffTauHe_L1} but for the L2 simulation.}
\label{EffTauHe_L2}
\end{figure}

In order to quantify the evolution in the transmissivity of the IGM and compare
our simulations against observations, we compute the effective optical depth
$\tau^{\tr{eff}}=-\ln \langle F(z)\rangle_\tr{los}$, where $F$ is the transmitted flux
($0\leq F \leq 1$) and the symbols $\langle \cdot \rangle_\tr{los}$ denote the average
over many lines of sight.
In Figs \ref{EffTauHe_L1} and \ref{EffTauHe_L2}, we show
the mean \heii\ effective optical depth, $\tau^{\tr{eff}}_{\tr{\heii}}$, evaluated from
$100$ random spectra in the L1, L1b and L2 simulations using spectral bins
with a size of $\Delta z=0.04$ as in \citet{Worseck}.
Data obtained from observations of high-redshift quasars \citep{Shull, Worseck}
and simulations \citep[][as computed in \citealp{Worseck} with bins $\Delta z=0.04$]{McQuinn} are also displayed for comparison.
For $z<3.2$, our results lie in the same ballpark as the experimental data.
At higher redshifts, not yet probed
by observations, the effective optical depth in the L1, L1b and L2 simulations
presents strong oscillations connected to the fluctuating number of ionizing
photons in the volumes.
Note that the skewness of the distribution of $\tau^{\tr{eff}}_{\tr{\heii}}$ implies that the
evaluation of the mean optical depth is mostly influenced by those regions of \heii\ spectra
with lower opacity therefore the mean is substantially different from the median.
Our results for $\tau^{\tr{eff}}_{\tr{\heii}}$ are in good agreement with
those presented by \citet{McQuinn} at low redshifts ($z<3$) especially
once the uncertainties of the models are taken into account,
but in our simulations $\tau^{\tr{eff}}_{\tr{\heii}}$ increases more gradually at
higher redshifts.
We do not show analogous results from the S boxes because
of the limited wavelength range covered by these simulations which corresponds to $\Delta z \simeq 0.02$.

\begin{figure}
\centering
\includegraphics[width=0.50\textwidth]{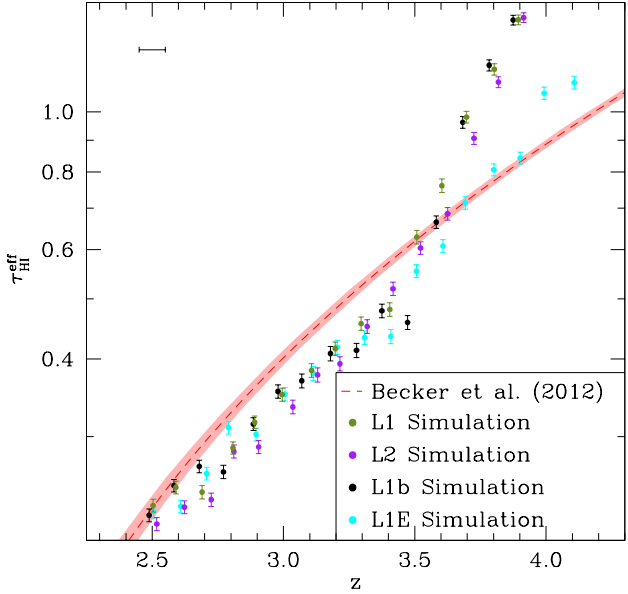}
\caption{Redshift evolution of the \hi\ effective optical depth in the L1 (green), L1b (black), L2 (purple) and L1E (cyan) simulations evaluated using bins of $\Delta z=0.1$ (as indicated by the horizontal error bar in the top-left corner of the plot).
The analytic fit to observational data (red dashed line) by \citet{BeckerTau} together with their $1\sigma$ uncertainty (shaded region) are also reported for comparison. Data from the L1b and L2 simulations have been slightly shifted in redshift ($-0.02$ and $+0.02$, respectively) to ease readability. Error bars indicate the standard error of the mean.}
\label{EffTauH}
\end{figure}

\begin{figure}
\centering
\includegraphics[width=0.5\textwidth]{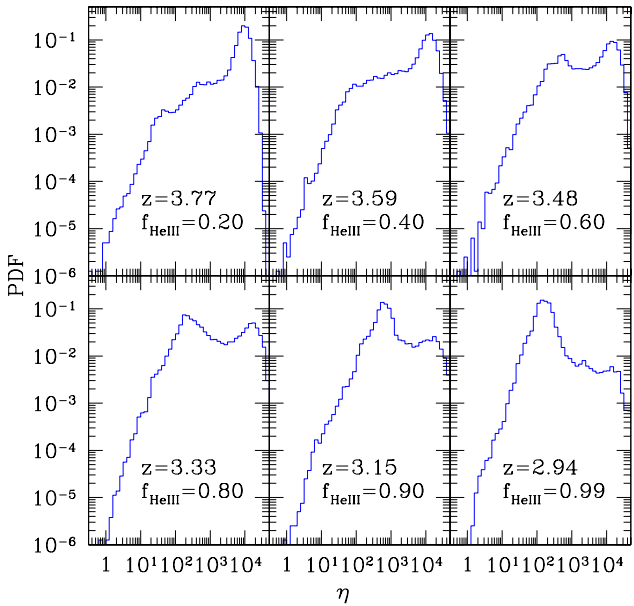}
\caption{PDF of the column-density ratio $\eta=\frac{N_\tr{\heii}}{N_\tr{\hi}}$ at different redshifts from $100$ lines of sight in the L1 simulation.}
\label{eta_distr}
\end{figure}

In Fig. \ref{EffTauH}, the \hi\ effective optical
depth, $\tau^{\tr{eff}}_{\tr{\hi}}$, evaluated from our L volumes
is compared with the most
recent measurements of the mean transmitted flux in the Ly$\alpha$ forest by
\citet{BeckerTau}:
\be
\tau^\tr{eff}_\tr{\hi}(z)=\tau_0\left(\frac{1+z}{1+z_0}\right)^\beta + C,
\ee
with $z_0=3.5$, $\tau_0=0.751$, $\beta=2.90$ and $C=-0.132$.
At $z>3.7$, the effective optical depth in the L1, L1b and L2 simulations is higher
than observed and shows a steep evolution. This is not
surprising since we start our radiative-transfer calculations in these volumes at $z_\tr{AGN}=4$ and
no AGN contribution to the UV background has been considered before. In fact, measurements of
the effective optical depth in the L1E simulation show a better agreement with the
observed $\tau^{\tr{eff}}_\tr{\hi}$ already at $z\simeq 3.9$.
For $2.5<z<3.7$, the effective
optical depth for all simulated volumes departs by less than 10 per cent from
the observational results even though it tends to be systematically lower.
This is remarkable as we do not need to
artificially rescale the \hi\ optical depth of our spectra to match observations
as often done in the literature.

\subsection{The \heii/\hi\ column-density ratio}

Absorption features in \hi\ and \heii\ spectra trace the same underlying
density field. Therefore, the ratio of the column densities
$\eta=N_\tr{\heii}/N_\tr{\hi}$ tracks the ratio of ionizing fluxes at $1$ and
$4\,\tr{ryd}$ and can be considered a measurement of the hardness of the UV
background.
Fig. \ref{eta_distr} presents the PDF of
$\eta$ along $100$ lines of sight at different redshifts in our L1 simulation.
Instead of fitting Voigt profiles to the mock spectra, the value of
$\eta$ is obtained directly from the ratio of the number densities
in each cell of the simulation. This way we can also probe optically thick regions.
At high redshift, the distribution is dominated
by a single peak at $\eta\simeq 10^4$, corresponding to spectra where the
\heii\ flux is highly absorbed while the \hi\ spectra show some transmission.
As \heii\ reionization progresses, a second peak
between $\eta \simeq 10^2$ and $10^3$
forms due to the increasing AGN contribution to the UV
background. When the reionization is close to completion ($z\simeq 2.9$), only the broad peak centred at
$\eta \simeq 100$ is present: by this time, the UV background is completely
dominated by the AGN contribution. Finally, at $z< 2.7$ (not shown in the figure) 
$\eta$ shows a small scatter around a central value of $\sim 50$ and is never larger than a few hundred.
Current observations are likely able to identify only the peak at low $\eta$.
In fact, complete absorption in \heii\ (within the noise level) precludes
precise measurements of very large values of $\eta$. Only lower limits
can be given in this situation.
Note that at high redshift, some cells in our simulations are characterized by a value of $\eta$
smaller than unity.

\subsection{Flux-transmission windows and dark gaps}
\label{ftw}

At high redshift,
the regions of transmitted flux in \heii\ spectra are relatively narrow in
wavelength and are clearly bounded by complete absorption.
In the following, we will refer to each of these regions as a
`Flux-Transmission Window' (FTW).
We operatively define an FTW as a region in the \heii\ spectrum where the flux is above $0.2$ times the continuum level.
Within the local Gunn--Peterson approximation, a $1$ per cent fraction of
singly ionized helium is enough to completely
saturate, with $\tau\gg 10$, the Ly$\alpha$ absorption, even in the
underdense regions where $\Delta_\tr{b}\simeq 0.1$.
Therefore, we expect to detect FTWs only if the medium is very highly ionized
and the recombination time is sufficiently long.
Our simulations reveal that, depending on redshift and
the extent of the ionization fronts, FTWs can occupy mildly overdense regions
surrounding AGNs or can be found in low-density environments, even
at large distances from any source of ionizing photons.

\begin{figure*}
\centering
\includegraphics[width=0.87\textwidth]{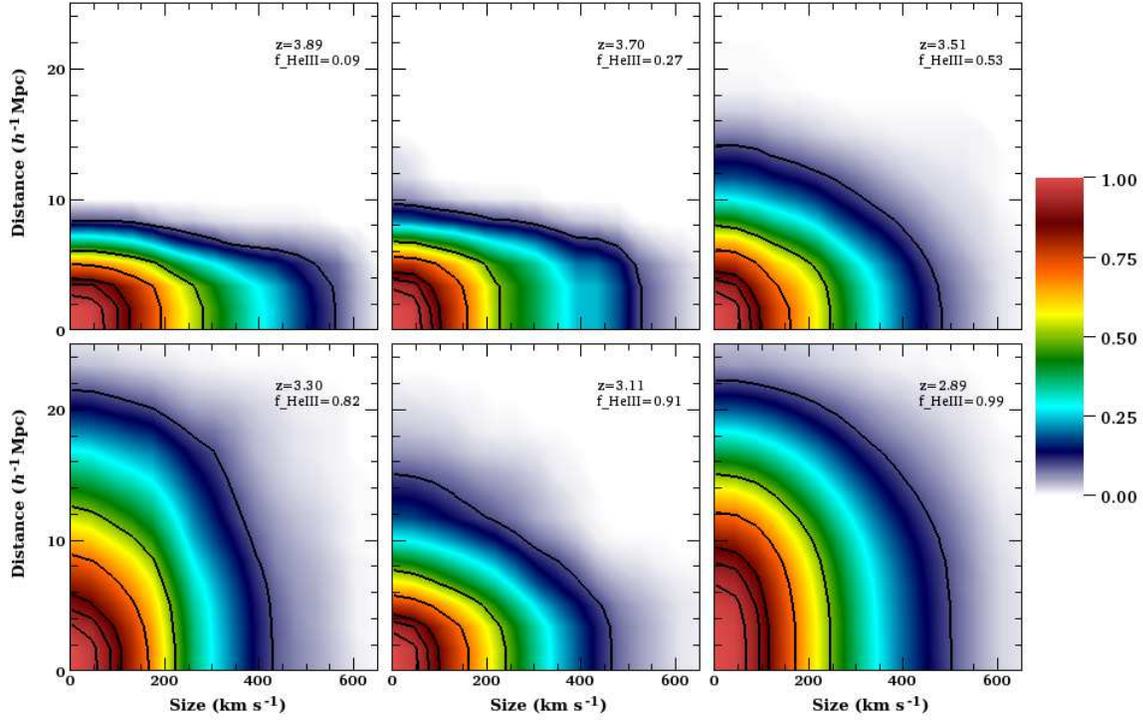}
\caption{Bivariate cumulative distribution function for the FTW size (in \kms) and the distance (in \mh) from the nearest source computed using $2500$ spectra extracted from the L1 simulation (PLE model). Contour levels enclose $5$, $10$, $15$, $30$, $50$ and $90$ per cent of the total number of points.}
\label{FTW_L1}
\end{figure*}

\begin{figure*}
\centering
\includegraphics[width=0.87\textwidth]{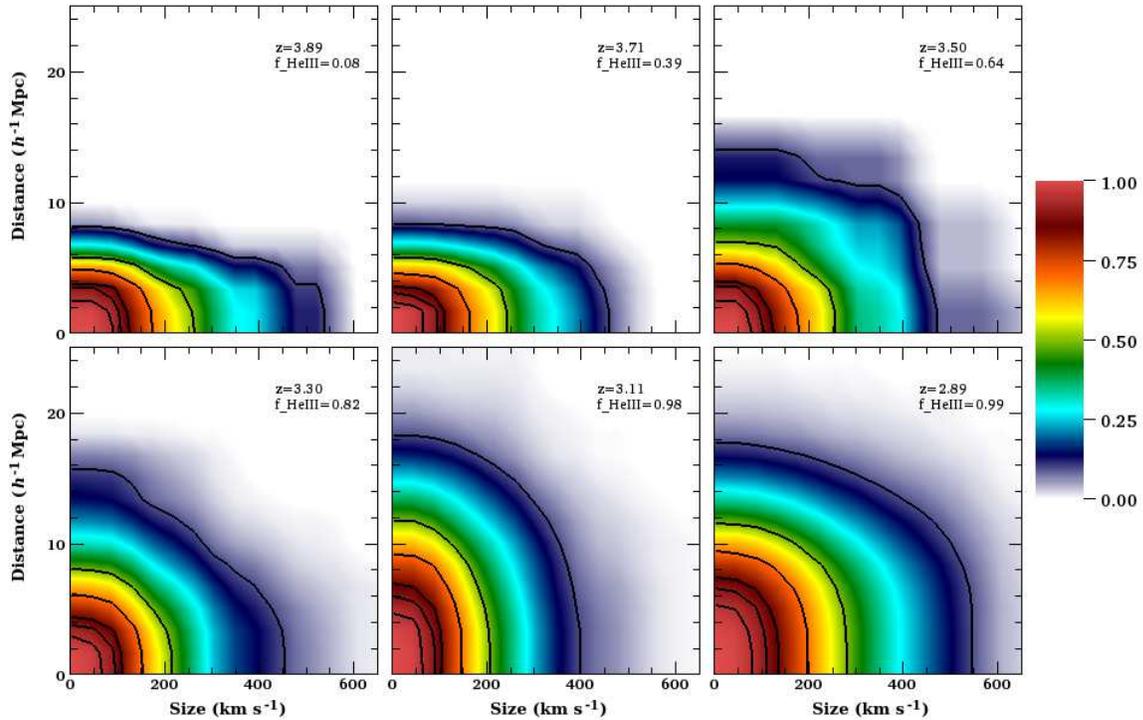}
\caption{As in Fig. \ref{FTW_L1} but for the L2 simulation (PDE model).}
\label{FTW_L2}
\end{figure*}

In Figs \ref{FTW_L1} and \ref{FTW_L2}, we present the cumulative joint
distribution of the FTW size (in \kms) and of the distance from the nearest
active source (in \mh) in the L1 and L2 simulations.
The colour of the pixel located at coordinates $(x,y)$ describes the
probability that an FTW has
size greater than $x$ and distance $y$ or higher from an active
source.
The PDE and PLE models produce very
similar results at all redshifts.
At the onset of \heii\ reionization, spectra from the L1 and L2 simulations
are characterized by extended ($\sim 300\,\tr{\kms}$) FTWs
at moderate distance from the ionizing sources. However, no FTWs appear at distances $>10\,\tr{\mh}$
from any AGN.
As reionization proceeds, the size of the transmission windows evolves only
marginally, while more and more FTWs can be found at large distances from the nearest
active source.
When \heii\ reionization is completed, FTWs typically extend for
$\sim 250\,\tr{\kms}$, with a characteristic distance from the
nearest active source of $\simeq 15\,\tr{\mh}$ for the L1 simulation
and $\simeq 12\,\tr{\mh}$ for the L2 volume.
A transient feature is seen in the L1 box around $z=3.1$ 
and is related to a negative fluctuation in the number of ionizing photons 
in the simulation (see Appendix \ref{PLEvsPDE}).
The properties of the FTWs strictly correlate with the spectral
characteristics of the active sources and do not depend on the past ionization history of the medium.
Only sources undergoing an active phase leave a detectable imprint on the
IGM, even during the last stages of the reionization, when the medium is expected to be less influenced
by the adopted AGN model.

\begin{figure}
\centering
\includegraphics[width=0.50\textwidth]{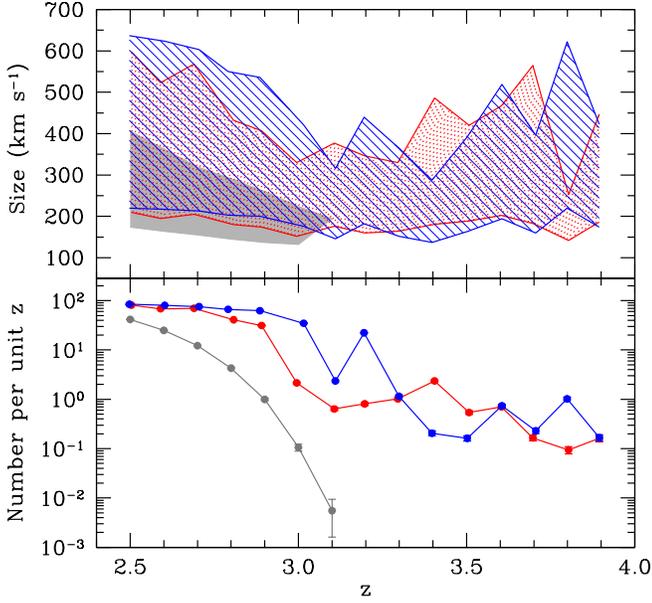}
\caption{Comparison between the FTWs in 2500 \heii\ spectra generated by different reionization scenarios.
Top: characteristic size of the FTWs
in the L1 (red), L2 (blue) and LHO (grey) simulations. The shaded area marks the region enclosed between the first and the third quartile. Bottom: number density of FTWs per unit redshift (error bars show the standard error of the mean). We neglect FTWs with size smaller than $88\,\tr{\kms}$, corresponding to the nominal resolution of our simulations. No FTWs are detected for 
$z > 3.1$ in the LHO box.}
\label{ftws}
\end{figure}

\begin{figure}
\centering
\includegraphics[width=0.50\textwidth]{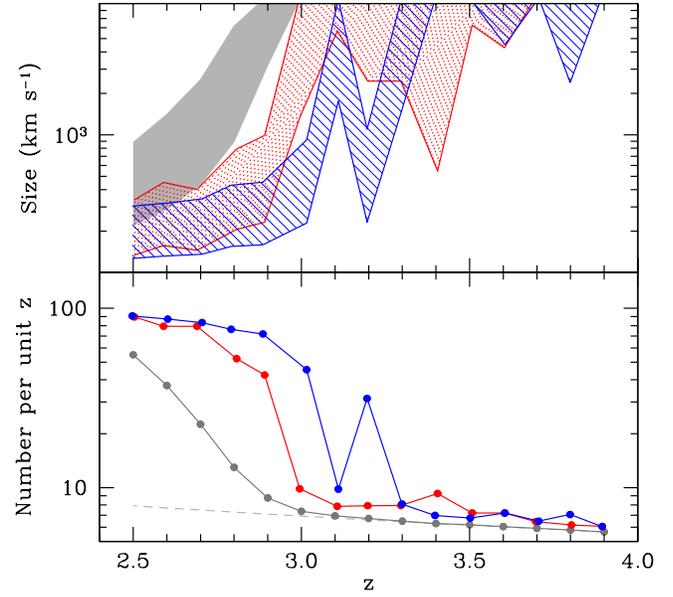}
\caption{As in Fig. \ref{ftws} but for the dark gaps. Note that the top panel shows the range between the median and the $10^\tr{th}$ percentile. The dashed line represents the minimum number of dark gaps that can be detected in our volume (one per line of sight).}
\label{darks}
\end{figure}

At redshift $z \gtrsim 3.5$, regions with a low optical depth in the Ly$\alpha$
forest are very rare. An alternative method to analyse the statistical
properties of \heii\ spectra is the distribution of dark gaps in the
transmitted flux. We operationally define a dark gap as a region in the \heii\ spectrum where the
flux is below 20 per cent of the continuum level.
Figs \ref{ftws} and \ref{darks} show the evolution with
redshift of FTWs and dark gaps in the L1 and L2
simulations, compared with the homogeneous reionization scenario provided by
the LHO simulation. Data are extracted from $2500$ spectra taken along random lines of sight.
The characteristic
size of the FTWs presents a similar evolution with redshift in the L1 and L2 simulations.
Minor differences can be related to fluctuations in the number of ionizing photons in the 
two volumes.
At $z\geq 2.8$, the number density of FTWs per unit redshift
comes out to be a promising statistic to investigate
the degree of patchiness of \heii\ reionization: in the
PLE and PDE models, this quantity is more than one order of magnitude larger than what is
expected for a reionization scenario with homogeneous UV background.
Note that for $z>3.1$ no FTWs are detected in the homogeneous-reionization run
where the \heii\ effective optical depth is slightly
higher than what is found with observations.

At later times, when \heii\ ionization is nearly
complete, both inhomogeneous models and the homogeneous reionization produce a comparable number of FTWs.
On the other hand, the number density of dark gaps (Fig. \ref{darks}) presents only small deviations
between the different reionization scenarios at high redshifts.
However, as it might be expected, the median length of the dark gaps in the PLE and PDE
scenarios drops to $\sim 1000$ \kms\ much earlier than in the homogeneous case. 
Note that the properties and the distribution in size
of FTWs and dark gaps are only marginally influenced by the adopted inhomogeneous
model: evolving with redshift the luminosity or the number density of the sources of ionizing
radiation does not produce significant differences in the FTWs and dark gaps statistics,
once random fluctuations in the number of ionizing photons are taken into 
account.

\begin{figure*}
\centering
\includegraphics[width=1\textwidth]{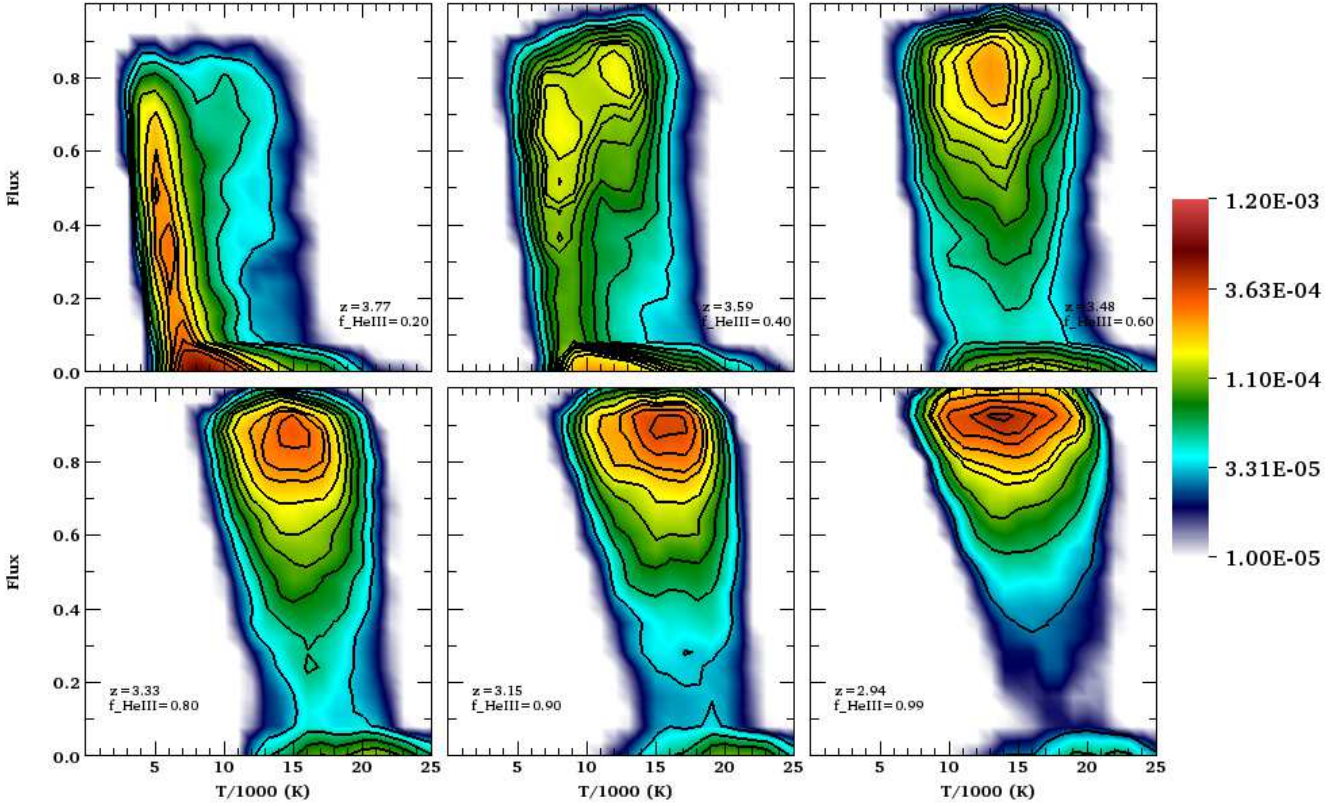}
\caption{Bivariate joint-probability density (in $\tr{K}^{-1}$) of the temperature and the corresponding \hi\ transmitted flux measured from $100$ simulated spectra in the L1 simulation.
Contour levels enclose from 10 to 90 per cent probability in steps of 10 per cent.}
\label{T-Flux}
\end{figure*}

\section[]{PROBING THE LOCAL IONIZATION HISTORY}

\subsection{Temperature bimodality} \label{Section:Tbimodality}

During the initial stages of \heii\ ionization, the IGM can be envisaged
as being composed of two phases (see Fig. \ref{phase_d_PLE}):
at fixed density, hot, already ionized
patches can be identified
around the locations where AGNs have been shining while
regions that have never been crossed by ionization fronts tend to be colder.
This temperature bimodality at fixed density leaves an imprint in quasar
spectra.
In Fig. \ref{T-Flux}, we present the joint probability distribution of
the temperature and the \hi\ transmitted flux (the matching is made at fixed
wavelength in the spectra as in Fig. \ref{spectrum1})
obtained considering $100$ random sightlines in the L1 simulation.
At the onset of helium reionization ($z\simeq 4$),
characteristic temperatures scatter around
$T\simeq 6 \times 10^3\,\tr{K}$ and the flux transmissivity is limited
to less than a few tens of per cent.
However, when the \heiii\ filling factor reaches 40 per cent or so, a strong
bimodal behaviour is established: high transmissivity tends to be associated
with high-temperature elements, while higher opacity corresponds to
lower temperatures.
At later times, when $f_{\tr{\heiii}}\geq 0.6$, the distribution becomes again unimodal and is characterized
by hot gas and relatively large transmissivity.
When the reionization is complete, the gas progressively cools down to lower temperatures.

\begin{figure}
\centering
\includegraphics[width=0.5\textwidth]{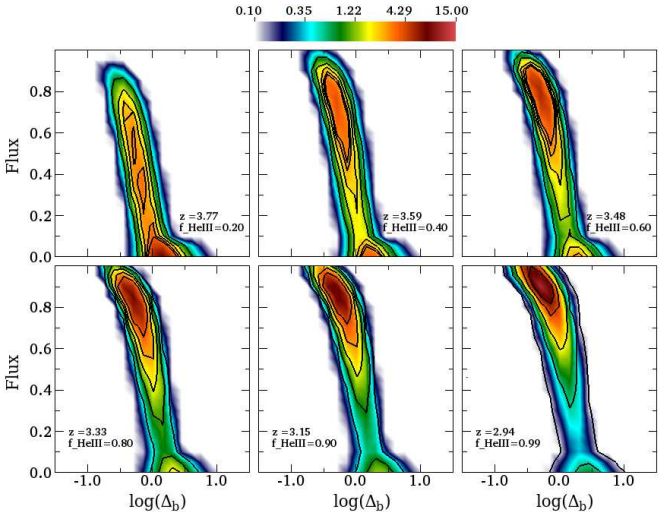}
\caption{Bivariate joint-probability density of the baryonic overdensity and the
\hi\ transmitted flux. Contour levels enclose $50$, $60$, $70$, $80$
and $90$ per cent probability.}
\label{Dens-Flux}
\end{figure}

It is not trivial to make a connection between Figs \ref{phase_d_PLE} and \ref{T-Flux}.
One needs to relate the gas density to the transmitted flux.
In Fig. \ref{Dens-Flux}, we show how the joint probability density of these quantities evolves
during \heii\ reionization. The peak of the distribution progressively shifts towards lower overdensities
and higher transmissivities with time. In other words, gas elements at fixed 
overdensity are associated with higher fluxes as reionization proceeds.
At fixed transmitted flux, \heii\ reionization shifts the characteristic overdensity
sampled by spectra towards higher $\Delta_\tr{b}$ and $T$.

\begin{figure}
\centering
\includegraphics[width=0.50\textwidth]{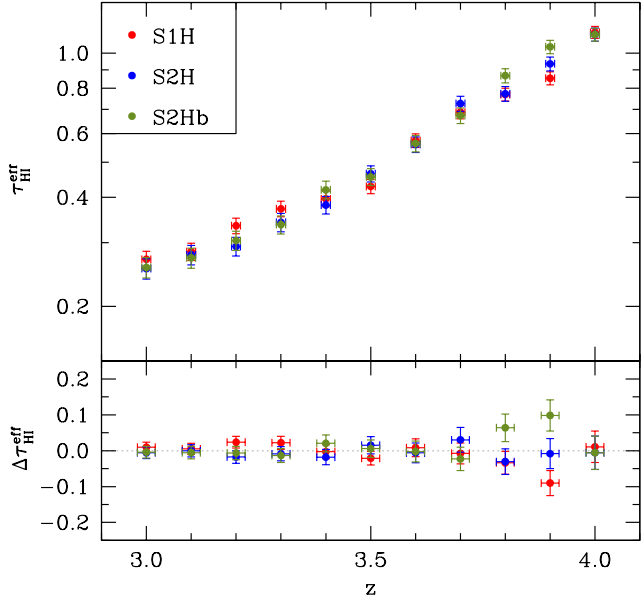}
\caption{Top: \hi\ effective optical depth in the S1H (red), S2H (blue) and S2Hb (green) volumes.
Data have been obtained from $150$ spectra using bins of $\Delta z=0.02$.
Error bars reflect the standard error of the mean flux.
Bottom: difference between the effective optical depth evaluated in the S1H, S2H and S2Hb simulations
and the mean of the three,
$\Delta\tau_\tr{\hi}^\tr{eff}=\tau_\tr{\hi}^{\tr{S}_\tr{i}}-\langle\tau_\tr{\hi}^\tr{eff}\rangle$.}
\label{S_comparison}
\end{figure}

\subsection{Observational probes of patchy reionization}

As a consequence of the patchiness of \heii\ reionization, widely separated volume elements get ionized at
slightly different times and this produces the above mentioned temperature bimodality of the IGM.
Can high-resolution quasar spectra distinguish
between \heiii\ regions and volumes that have yet to be ionized?
As an example, let us consider the two high-resolution simulations
that differ most in the ionization history: in the S1H box,
the volume filling factor of \heiii\ reaches 20 per cent
nearly $185\,\tr{\Myr}$ earlier than in the S2Hb volume (corresponding to $\Delta z\sim 0.35$,
see Fig. \ref{reion_trend}).
In Fig. \ref{S_comparison}, we compare the evolution of
$\tau_\tr{\hi}^\tr{eff}$ in the high-resolution boxes.
The effective optical depth in the S1H simulation drops very rapidly when the volume is
reached by ionization fronts and shows a milder evolution at later times.
When this box is already substantially ionized at $z\simeq 3.9$, ionization fronts have
yet to reach the S2Hb volume while the S2H simulation has an intermediate ionization state.
At this point, the scatter in $\tau_\tr{\hi}^\tr{eff}$
reaches a maximum. As reionization proceeds in all simulations, the evolution of the effective
optical depth in the three volumes is very similar ($z \leq 3.7$), with small fluctuations related to the
different properties of the ionizing sources in the boxes.
All this suggests that an increased scatter in transmissivity between different lines of sight
can be used as an indicator of the onset of \heii\ reionization.

In the remainder of this section, we investigate how a patchy reionization process and the temperature
bimodality of the IGM make an impact on to different statistical probes
that are commonly used to analyse the \hi\ Ly$\alpha$ forest in quasar spectra.
In order to resolve the scales associated with these absorption features we
mainly focus on the high-resolution simulations S1H and S2Hb.

\subsubsection{Distribution of the Doppler $b$ parameters}

\begin{figure*}
\centering
\includegraphics[width=1.\textwidth]{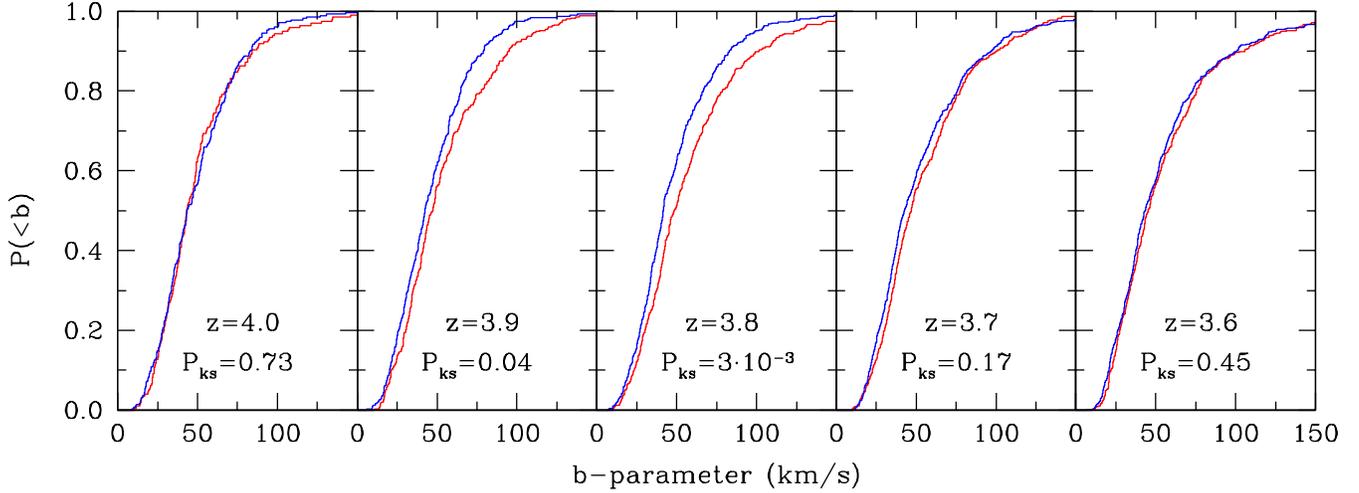}
\caption{Cumulative distributions of the Doppler $b$ parameters for Ly$\alpha$ lines with \hi\
column density $\log(N_{\tr{\hi}})<14.0\,\tr{cm}^{-2}$ and statistical error $\Delta b<10\,\tr{\kms}$.
Data extracted from $150$ spectra in the S1H and S2Hb volumes are plotted in red and blue, respectively.
The probability that the two samples
are extracted from the same parent distribution according to the Kolmogorov--Smirnov
test is given in each panel together with the corresponding redshift.}
\label{cumulative}
\end{figure*}

Using the {\sevensize VPFIT}\footnote{http://www.ast.cam.ac.uk/$\sim$rfc/vpfit.html} package,
we fit multiple Voigt profiles to the mock \hi\ absorption-line spectra extracted from the S1H and S2Hb
simulations.
In Fig. \ref{cumulative}, we compare the cumulative distributions of the Doppler $b$ parameters measured
in the two simulations. Only absorption lines with low column density, $\log{N_\tr{\hi}} < 14.0\,\tr{cm}^{-2}$,
and a small error on the $b$ parameter, $\Delta b < 10\,\tr{\kms}$, are considered.
We report in the figure also the probability that the two samples are extracted from the same
parent distribution according to the Kolmogorov--Smirnov statistics ($P_{\rm ks}$).
The difference between the distributions becomes statistically significant (at the $95$ per cent confidence
level) at $z=3.9$ and increases even further at $z=3.8$,
when the gas in the S1H box is being heated by the ionization fronts while
the \heiii\ filling factor in the S2Hb volume is still smaller than a few per cent.
The maximum discrepancy between the cumulative distributions is recorded for $b\simeq 56\,\tr{\kms}$.
Later on, as \heii\ reionization proceeds also in the S2Hb volume,
the distributions become more and more similar.
This analysis shows that the Doppler parameter of lines with low column density is very sensitive
to the temperature increase associated with the initial stages of the reionization process.
During the initial phases of \heii\ reionization the mean and the standard deviation of the $b$ parameter
distribution move from $50.5 \pm 29.2\,\tr{\kms}$ (with a median of $\simeq 43.2\,\tr{\kms}$) to $55.8 \pm 37.7\,\tr{\kms}$
(with a median of $\simeq 45.1\,\tr{\kms}$) as seen in observations for higher column densities \citep[][]{Rauch}.

\subsubsection{The curvature statistic}
\label{Section:curvature}

\begin{figure}
\centering
\includegraphics[width=0.50\textwidth]{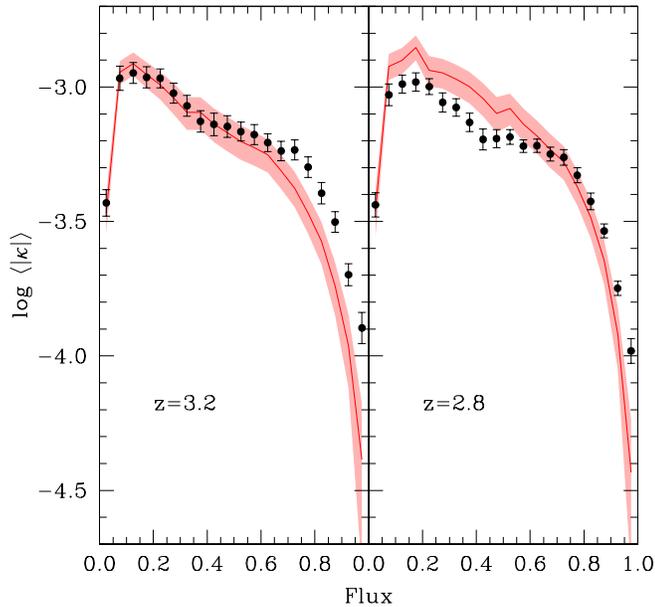}
\caption{Distribution of the mean absolute curvature as a function of the normalized flux near the end of \heii\ reionization in the S1H simulation.
The solid lines and the shaded regions show the values extracted from $450$ spectra and $2\sigma$ error bars, respectively. 
Points with error bars ($2\sigma$) are observational data from \citet{Curvature}.}
\label{curvature_data}
\end{figure}

\begin{figure}
\centering
\includegraphics[width=0.5\textwidth]{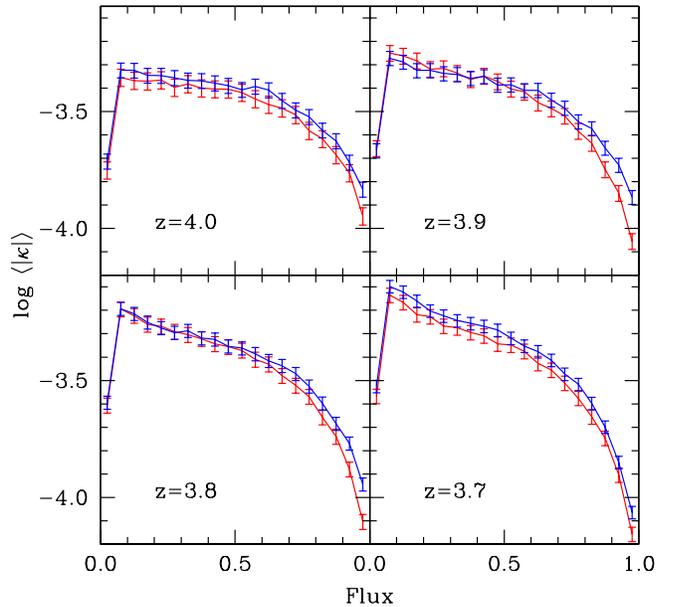}
\caption{Evolution with redshift of the mean absolute curvature as a function of the normalized flux.
The curvature is evaluated using $450$ spectra extracted from the S1H (red) and the S2Hb (blue) simulations.
Error bars show the standard error of the mean.}
\label{curvature_evol}
\end{figure}

An alternative method that has been proposed to measure the imprint of \heii\ ionization on the IGM is the
curvature statistic. This is defined in terms of the transmitted  \hi\ flux,
$F_\tr{\hi}$, as \citep{Curvature}
\be
\kappa \equiv \frac{F''_\tr{\hi}}{\left[ 1 + (F'_\tr{\hi})^{2} \right]^{3/2}}\,,
\label{curvdef}
\ee
where the prime denotes differentiation with respect to the velocity separation in the spectrum.
The curvature essentially coincides with $F''_\tr{\hi}$
as the denominator in Equation (\ref{curvdef}) is always very close to unity.
This statistic does not require the decomposition of the Ly$\alpha$ forest into single
absorption features and is thus suitable also when the absorption lines are strongly
blended.
We consider $450$ mock absorption spectra extracted from the S1H and S2Hb simulations.
In order to measure the curvature, we first fit each spectrum with a cubic spline
and normalize the flux to the maximum value of the fit.
The evolution with redshift of the mean absolute curvature, $\langle|\kappa|\rangle$,
reproduces the trend seen in observational data \citep{Curvature} for $z<3.5$, as shown in Fig. \ref{curvature_data}
for the S1H volume.
In Fig. \ref{curvature_evol}, we compare the evolution with redshift of the mean 
absolute curvature evaluated in the S1H and in the S2Hb simulations.
As soon as the \heii\ ionization fronts have swept a small fraction of the volume,
the mean absolute curvature in the S1H box departs from the results in the S2Hb simulation,
when considering bins of high transmissivity.
Later on, when the local \heiii\ filling factor in the S2Hb box
increases ($z \leq 3.7$), the two distributions are again very similar.
Although it is a small effect, the redshift evolution of  $\langle|\kappa|\rangle$
might be able to identify the onset of \heii\ reionization.
Probing different lines of sight, one should be able to
distinguish regions with different ionization histories by measuring the mean absolute curvature in narrow redshift bins
of high transmissivity. For instance, we expect that the temperature bimodality which arises in our models at $3.6<z<4.0$
should be reflected in a broad and possibly bimodal distribution of the mean absolute curvature in real data.

\section[]{CONCLUSIONS}

We have investigated the epoch of \heii\ reionization using
a suite of AMR hydrodynamic simulations with different box sizes.
AGNs with a lifetime of $45\,\tr{\Myr}$ have been associated with dark matter haloes using
a Monte Carlo method designed to reproduce the observed luminosity function at
redshift $z=4$. For simplicity,
the change in the AGN emissivity with time has been modelled
assuming either pure density or pure luminosity evolution of the sources.
Numerical radiative transfer of the ionizing radiation emitted by AGNs has been
carried out in post-processing through a snapshot of the simulations.
The UV emission by galaxies has been represented with a time-evolving but
spatially uniform background.

We have studied how \heii\ reionization alters the thermal state
of the IGM as well as its impact on the \hi\ and \heii\
Ly$\alpha$ forest. Several statistics have been employed to probe into
the topology of the reionization process in the redshift range $2.5<z<4$.
Our main results can be summarized as follows.

\begin{enumerate}
\item In agreement with observations \citep{Shull, Worseck},
we find that
\heii\ reionization is a process extended in redshift ($\Delta z \gtrsim 1$)
characterized by a large sightline-to-sightline variance in \heii\ absorption.
Well-separated
ionized bubbles inflate around the location of the hard UV sources
and start recombining when AGN activity terminates.
Over time, bubbles percolate and the complete ionization of the IGM takes place.
The \heiii\ fraction and the temperature of the ionized IGM strongly correlate
with the properties (luminosity and hardness) of the nearest AGNs,
quickly losing memory of the previous ionization history of the gas.
Within a single bubble, the gas properties are heterogeneous and
depend on the underlying density distribution with a tight temperature--density
relation. Small neutral islands form due to self-shielding of
high-density regions which ionize last and reach the highest temperatures.
The degree of patchiness in the \heiii\ distribution
is affected by the lifetime of the sources and the properties
(number density and luminosity distribution) of the AGN population.
On the other hand,
the overall duration of the reionization epoch is rather model independent
and only depends on the total emissivity of the UV sources.

\item The evolution of the \hi\ and \heii\ effective optical depths in our simulations
match data from high-resolution observations
\citep[e.g.][]{Shull, Worseck, BeckerTau}.
The \heii\ to \hi\ column-density ratio, $\eta$, shows a strong spatial
variability during \heii\ reionization
and assumes values ranging between $0.5$ and $5 \times 10^4$.
Its PDF evolves with redshift reflecting
the increasing volume fraction in which the UV radiation field is harder
due to the AGN emission. The contribution of the ionized regions creates
a peak at $\eta\simeq 50-100$ which becomes more and more prominent
between $2.5<z<4$. This peak should be detectable from real data
once a significant number of \heii\ quasars is available for spectral studies.
In our simulations,
the \hi\ effective optical depth evolves smoothly during \heii\ reionization.
In agreement with previous work \citep{Bolton2009a, Bolton2009b, McQuinn},
we do not recognize any sharp feature reminding of the
narrow dip observed at $z\simeq 3.2$ \citep{Theuns,Bernardi,DallAglio,Faucher}.
This is likely a consequence of the long duration of the reionization process.

\item The number density of the flux-transmission windows and
the mean size of the dark gaps in the \heii\ spectra
provide a powerful diagnostic tool to distinguish between different reionization
scenarios (e.g. homogeneous versus inhomogeneous). Albeit these statistics encode information 
about the luminosity, number density and clustering properties of the sources 
responsible for reionization, they cannot discriminate between our PDE and PLE models.
Fluctuations in the transmitted flux correlate with the local intensity
of ionizing radiation.
Very extended regions of transmitted flux are generally located next to bright
AGNs even though there are rarer cases where flux-transmission windows with
moderate size (100 \kms) can lie up to 22 \mh\ away from the nearest AGN.

\item The IGM is heated during \heii\ reionization:
gas at mean density experiences a temperature jump of $\sim 12\tn000\,\tr{K}$ and
cools afterwards.
We find that the typical temperature increment from $z=4$ to $3$
(at mean density) ranges between $9000$ and $10\tn000\,\tr{K}$.
This value is in good agreement with
results based on the curvature of \hi\ Ly$\alpha$ spectra \citep{Curvature}.

\item The patchiness of the \heii\ reionization process gives rise to a bimodal
distribution of the IGM temperature.
Gas elements populate two distinct sequences in the temperature--density plane
depending on their \heiii\ fraction. At fixed density, ionized gas is hotter.
The scatter around these effective equations of state is much smaller than the
separation between the sequences. This is a transient feature
which is particularly evident when $f_\tr{\heiii}\sim 0.4$.
As reionization proceeds,
more and more gas moves from the cold to the hot sequence.
When \heii\ reionization is completed, only a single and well-defined
temperature--density relation is present.
At $z\simeq 2.9$, this is
characterized by a polytropic index $\gamma\simeq 1.20$ which is
lower than the corresponding value at the onset of reionization
($\gamma\simeq 1.56$).
The average temperature at mean density is
$T_0\simeq 1.56 \times 10^4$ K, in agreement with observations \citep{Curvature} but
slightly lower than in previous theoretical work \citep{McQuinn, Meiksin}.
We find no evidence for an inverted equation of state of the IGM at low densities $\Delta_{\rm b}<5$.

\item As a consequence of the bimodality in the temperature,
neutral and ionized patches of the IGM are characterized by different
probability distributions of the Doppler $b$ parameters and of the
curvature of \hi\ spectra.
Therefore, we predict that \heiii\ regions should generate \hi\ spectra with
higher $b$ parameters and a reduced mean absolute curvature in bins 
characterized by high transmissivity.
These statistics are particularly sensitive during
the initial stages of the reionization process,
when the \heiii\ volume filling factor is not too large.
\end{enumerate}

This work presents a detailed theoretical study of the epoch of \heii\
reionization. Our analysis is based on the current understanding of the subject
and uses the most advanced numerical tools.
Many details of the model, however, are still prone to improvement.
The properties of the sources of radiation and their time evolution,
for instance, are still described with a toy model. Radiation transport
is decoupled from hydrodynamic processes.
In spite of this, our simulations provide a realistic description
of the history of reionization which is compatible with all observational
data. Our results, albeit more qualitative than quantitative, provide
a starting point for novel investigations of the IGM. For instance, it
would be interesting to analyse existing data on the \hi\ forest at $3<z<4$
in light of the temperature bimodality seen in our simulations.
Times are mature for investing in this challenge.
In the next few years, studies of new sightlines towards \heii\ quasars
will provide tighter constraints on many observables.
At the same time,
hydrodynamical simulations of cosmological volumes with increasing spatial
resolution and more sophisticated radiative transfer (hopefully fully coupled
to hydrodynamics) will become available.
The reciprocal feedback between theory and observations will be key to
shed new light on the physics of \heii\ reionization.

\section*{ACKNOWLEDGEMENTS}
We are grateful to Romain Teyssier for help with the {\sevensize RAMSES} code and
Francesco Haardt for providing a model for the galaxy UV background.
MC thanks Gabor Worseck, Xavier Prochaska and Joseph Hennawi for useful
comments on an earlier version of the paper and
Patrick McDonald for advice on the flux power spectrum.
We thank the anonymous referee for comments that improved the quality of our paper.
We acknowledge use of the visualization software {\sevensize VISIT} developed
by the US Department of Energy Advanced Simulation and Computing Initiative.
The hydrodynamical AMR simulations were run at the Leibniz-Rechenzentrum in
Munich.
This work was supported by the Deutsche Forschungsgemeinschaft (DFG)
through the project SFB 956 \ti{Conditions and Impact of Star Formation}.
SC acknowledges support from the NSF grant AST-1010004.

\appendix

\section[]{Hydrodynamic Response}
\label{hydro}

Our study is based on numerical simulations post-processed with a radiative-transfer code and thus neglects
the hydrodynamical response of the gas to the heating due to \heii\ reionization. In order to estimate
the accuracy of our approximations, we compare the output of two periodic AMR simulations with identical initial conditions.
The first box (T1) is illuminated with the UV background radiation computed as in \citet{HM2012} considering only the
galaxy contribution. In the second one (T2), instead, an extra contribution from AGNs is turned on for $z<4$.
In both cases, the computational volume has a linear size of $25\,\tr{\mh}$
and the mesh structure matches the spatial resolution of our L simulations.

\begin{figure}
\centering
\includegraphics[width=0.5\textwidth]{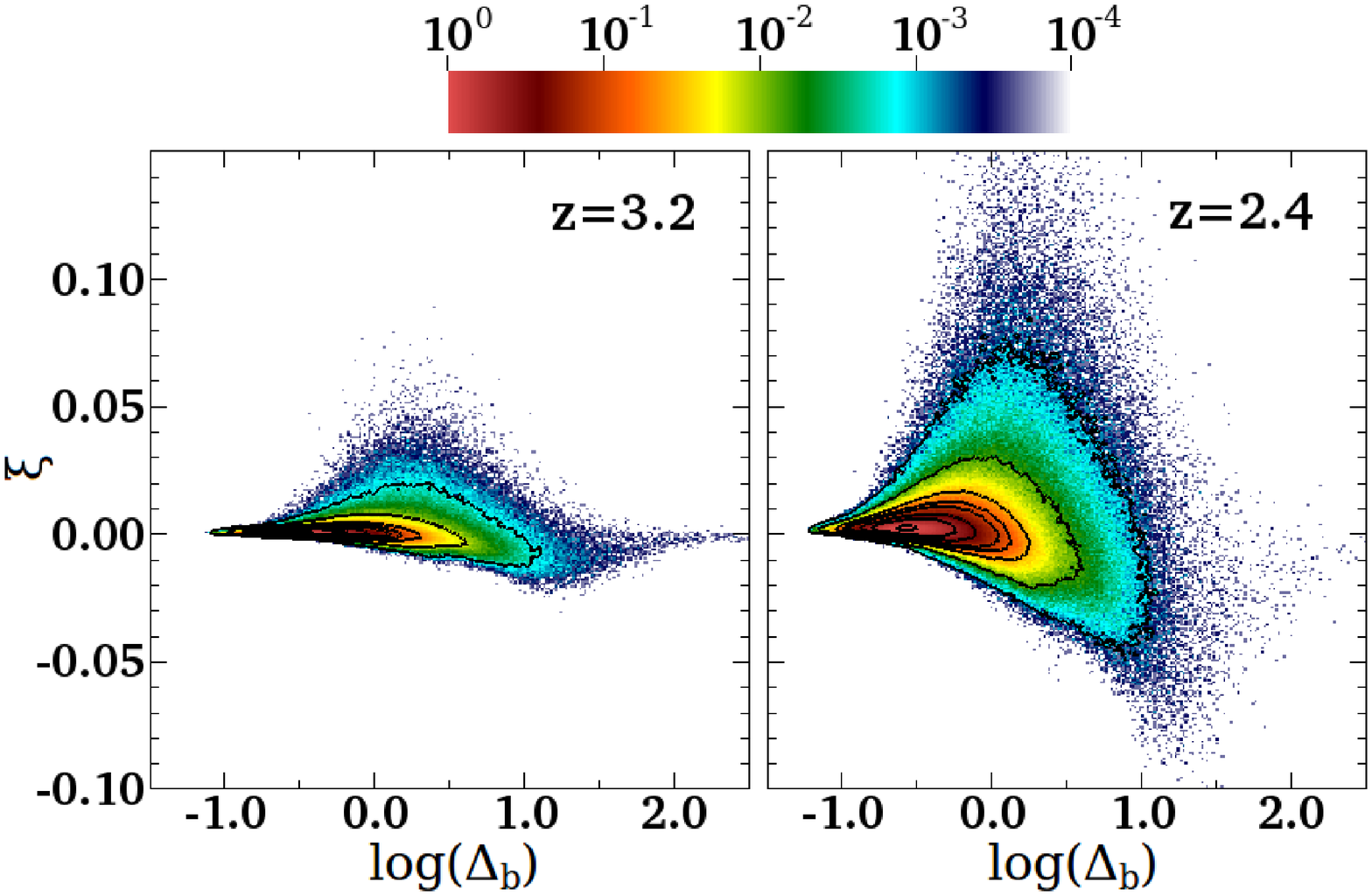}
\caption{Hydrodynamical response of the gas to \heii\ reionization. Joint distribution of the gas overdensity ($\Delta_{\rm b}$) and 
the relative change of the local gas density between the T2 and T1 simulations ($\xi$). The color scale indicates the gas mass in each pixel (normalized to unity at maximum).
Contour levels correspond to $0.1$, $1$, $5$, $10$, $15$, $30$, $50$ and $90$ per cent of the maximum value.}
\label{hydro_resp1}
\end{figure}

\begin{figure}
\centering
\includegraphics[width=0.5\textwidth]{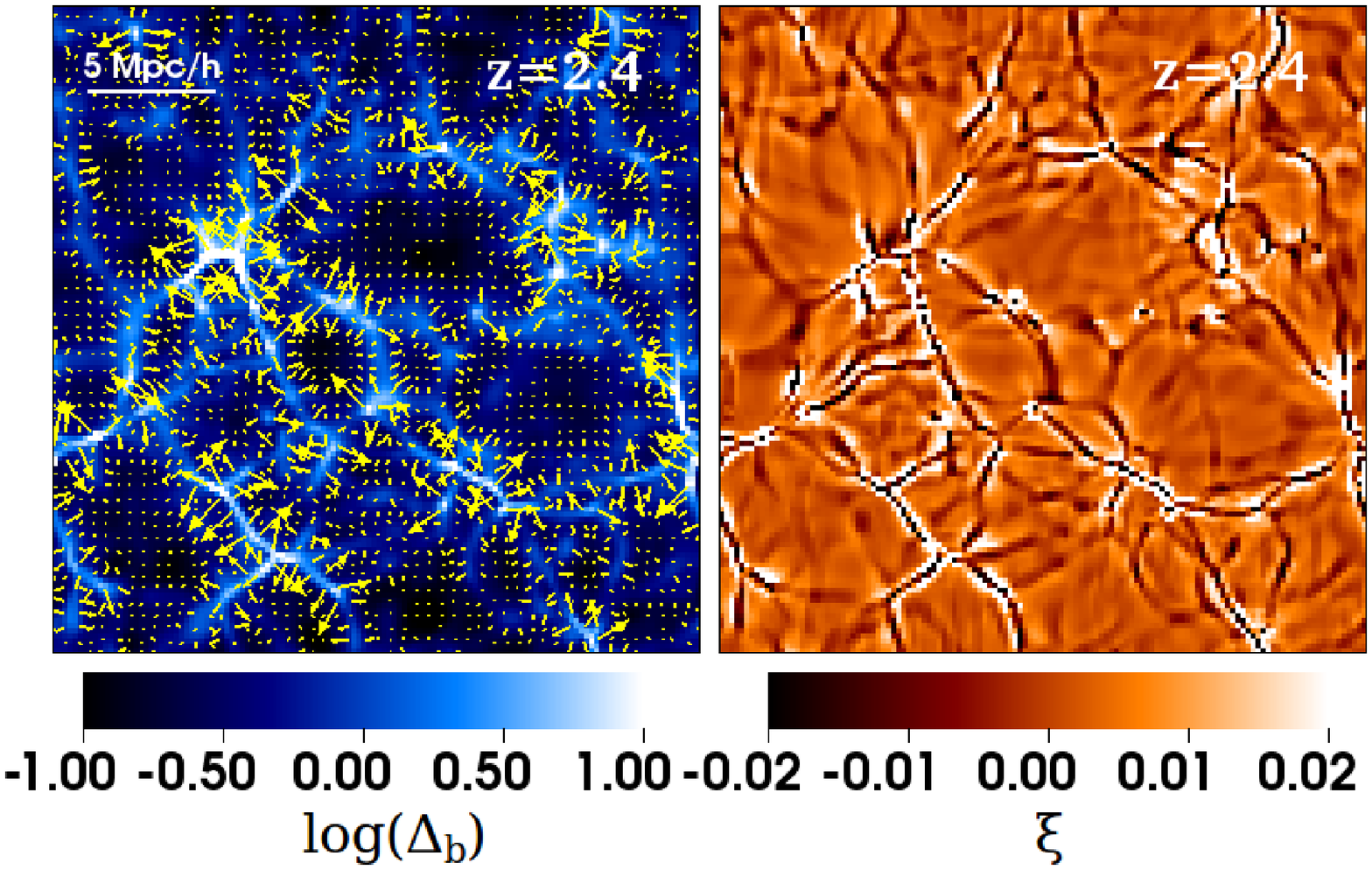}
\caption{Left: density distribution in a slice of the T2 simulation. The transverse size is of $25\,\tr{\mh}$ and
the yellow arrows indicate the velocity difference between cells in the T2 and in the T1 simulations. Right: spatial distribution of the variable $\xi$ in the same slice.}
\label{hydro_resp2}
\end{figure}

The distribution of the relative difference between the densities at the same spatial location in the two simulations,
\be
\xi(\boldsymbol{x},z)=\frac{\Delta_{\rm b}^\tr{T2}(\boldsymbol{x},z) - \Delta_{\rm b}^\tr{T1}(\boldsymbol{x},z)} {\Delta_{\rm b}^\tr{T2}(\boldsymbol{x},z)},
\ee
is shown in Fig. \ref{hydro_resp1} for two different redshifts. Overdense regions with $\Delta_{\rm b}
\sim 10$ get less dense during \heii\ reionization while underdense regions with $\Delta_{\rm b}\sim 0.1$ are
basically unaffected by it. The regions around mean density are the most altered by \heii\ reionization: they get
denser and denser with time.
A detailed inspection of the velocity and density changes (see Fig. \ref{hydro_resp2}) reveals that
during \heii\ reionization
dense filaments are heated and expand thus making their central regions thinner
and their outer layer denser. Volume-filling voids experience minor changes.
For the vast majority of the gas, neglecting the hydrodynamical response produces
an error in the density which is smaller than $8$ per cent by $z=2.4$.
Our results are in line with the parallel analysis by \citet{Meiksin}.

\section[]{Number of spectral bins} \label{convergence_test}

\begin{table}
\centering
\begin{tabular}{l c c c}
\hline\hline
 & \hi\ & \hei\ & \heii\ \\ [0.5ex]
\hline
Minimum energy (ryd) & 1      & 1.8088 & 4  \\
Maximum energy (ryd) & 1.8088 & 4      & 40 \\
\hline
No. of spectral bins:\\
~~Run 1 & 5  & 5  & 10 \\
~~Run 2 & 7  & 7  & 22 \\
~~Run 3 & 10 & 10 & 30 \\
~~Run 4 & 15 & 15 & 45 \\
~~Run 5 & 20 & 20 & 60 \\[1ex]
\hline
\end{tabular}
\caption{Number of spectral bins used to test the convergence of the radiative-transfer scheme.}
\label{table:spectral_bins}
\end{table}

\begin{figure}
\centering
\includegraphics[width=0.5\textwidth]{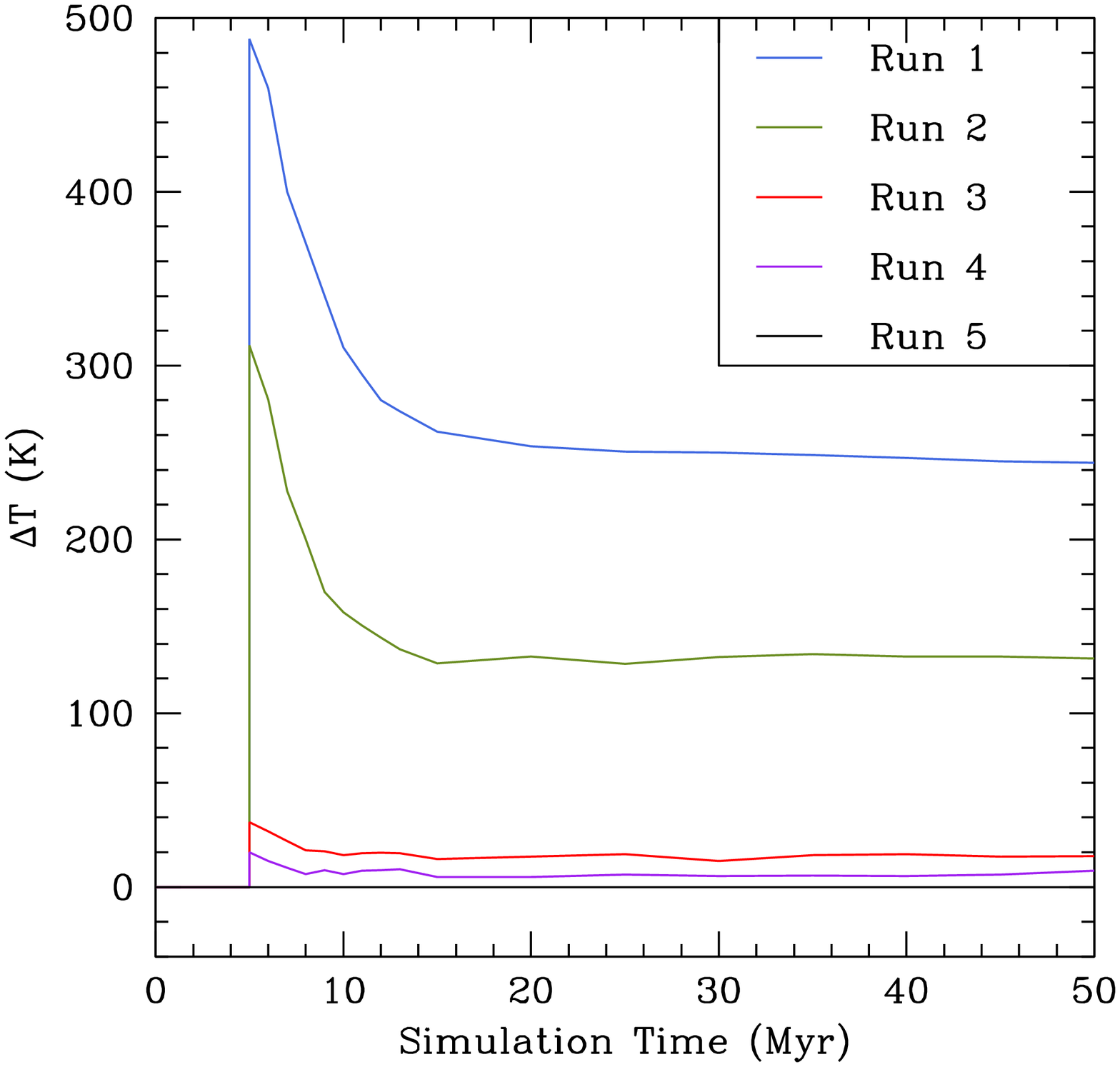}
\caption{Influence of the frequency binning in the radiative-transfer scheme on to the temperature of a cell at mean density
(see the main text for details). The variable $\Delta T$ indicates the difference between the actual temperature
in a simulation and what is measured in the reference case, Run 5.
Each curve corresponds to one of the simulations in Table \ref{table:spectral_bins} as indicated by the label.}
\label{spectral_bins}
\end{figure}

{\sevensize RADAMESH} samples the UV spectrum of the ionizing sources with a finite number of frequency bins which
are logarithmically spaced.
In order to optimize this choice,
we perform a convergence study by considering a simulation box with a linear size of $25\,\tr{\mh}$.
We start with a base grid of $64^3$ elements and add four levels of refinement during the evolution.
The box is uniformly illuminated with a soft UV background generated by galaxies.
At $z=3$, a single AGN (with a steady magnitude of $M_{1450}=-20.3$ and located at the centre of a massive
dark matter halo) is turned on. We run a series of simulations by varying the binning strategy as detailed in
Table \ref{table:spectral_bins}. For each simulation, we follow the propagation of radiation for $50\,\tr{Myr}$.
To illustrate our results, we focus on a cell at mean density which is located
$\simeq 1.1\,\tr{\mh}$ away from the active AGN.

In Fig. \ref{spectral_bins}, we show the evolution of the temperature difference between each run and Run 5 which
uses the largest number of spectral bins.
Using too few bins overestimates the temperature of the gas, especially during the passage of the ionization front.
For the number of spectral bins adopted in this work, the temperature increment with respect to the reference case is
negligible.

Finally, it is interesting to quantify the impact of soft X-rays with energies above $40\,\tr{ryd}$ that have been
neglected in our study. For this reason, we run an additional simulation in which we extend the maximum photon energy to
$80\,\tr{ryd}$ and partition the \heii\ ionizing photons into 60 spectral bins.
The extra heating caused by the energetic photons produces an early temperature increment of $\approx 150\,\tr{K}$
(due to their long mean free path) which reduces to less than $50\,\tr{K}$ a few \Myr\ after the passage of
the ionization front. We conclude that considering only photons with energies below $40\,\tr{ryd}$ does not
affect our results.

\section[]{Comparison between PLE and PDE source models} \label{PLEvsPDE}

\begin{figure}
\centering
\includegraphics[width=0.5\textwidth]{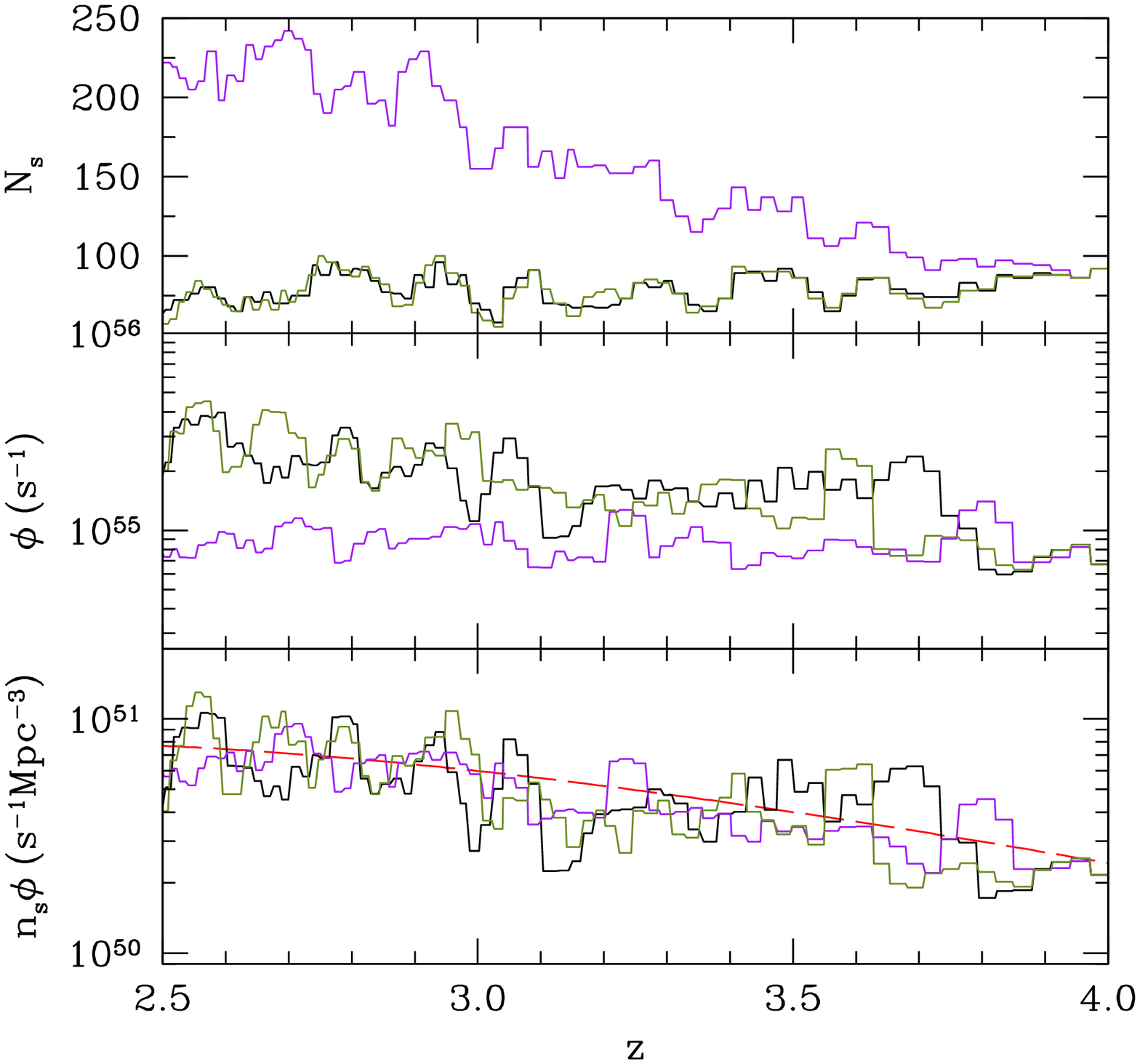}
\caption{Comparison between different AGN models used in the L1 (green), L1b (black) and L2 (purple) simulations.
Top: number of active sources. Centre: mean number of \hi\ ionizing photons emitted by a single source per
unit time. Bottom: total number of ionizing photons emitted per unit time and volume. The red dashed line
shows the model predictions by \citet{HM2012} that we use to calibrate
the time evolution of the AGN emissivity.}
\label{PLEvsPDE_evol}
\end{figure}

As discussed in Section \ref{Section:AGN_model}, we model the redshift evolution of the AGN emissivity either by modifying the number density of the sources (PDE model) or by altering their luminosity (PLE model).
For illustrative purposes and to facilitate understanding, we show in Fig. \ref{PLEvsPDE_evol} how the number
of active sources (top panel), the mean number of ionizing photons (with energy above $1\,\tr{ryd}$)
emitted by a single source per unit time (middle panel) and the total emissivity of ionizing photons per unit
time and volume (bottom panel) evolve in the L1, L1b and L2 simulations.

\bsp

\label{lastpage}


\begin{thebibliography}{99}
\bibitem[Becker et al.(2011)]{Curvature} Becker, G.~D., Bolton, J.~S., Haehnelt, M.~G., \& Sargent, W.~L.~W.\ 2011, \mnras, 410, 1096
\bibitem[Becker et al.(2012)]{BeckerTau} Becker, G.~D., Hewett, P.~C., Worseck, G., \& Prochaska, J.~X.\ 2012, arXiv:1208.2584
\bibitem[Bernardi et al.(2003)]{Bernardi} Bernardi, M., Sheth, R.~K., SubbaRao, M., et al.\ 2003, \aj, 125, 32
\bibitem[Bertschinger(2001)]{Bertschinger} Bertschinger, E.\ 2001, \apjs, 137, 1
\bibitem[Bolton et al.(2008)]{Bolton2008} Bolton, J.~S., Viel, M., Kim, T.-S., Haehnelt, M.~G., \& Carswell, R.~F.\ 2008, \mnras, 386, 1131
\bibitem[Bolton et al.(2009a)]{Bolton2009a} Bolton, J.~S., Oh, S.~P., \& Furlanetto, S.~R.\ 2009a, \mnras, 395, 736
\bibitem[Bolton et al.(2009b)]{Bolton2009b} Bolton, J.~S., Oh, S.~P., \& Furlanetto, S.~R.\ 2009b, \mnras, 396, 2405
\bibitem[Bolton et al.(2012)]{BoltonLifetime} Bolton, J.~S., Becker, G.~D., Raskutti, S., et al.\ 2012, \mnras, 419, 2880 
\bibitem[Bryan \& Machacek (2000)]{Bryan} Bryan, G.~L., \& Machacek, M.~E.\ 2000, \apj, 534, 57
\bibitem[Calura et al.(2012)]{Calura} Calura, F., Tescari, E., D'Odorico, V., et al.\ 2012, \mnras, 422, 3019
\bibitem[Cantalupo et al. (2007)]{Fluorescence07} Cantalupo, S., Lilly, S.~J., \& Porciani, C.\ 2007, \apj, 657, 135
\bibitem[Cantalupo \& Porciani(2011)]{Radamesh} Cantalupo, S., \& Porciani, C.\ 2011, \mnras, 411, 1678
\bibitem[Croft et al.(2002)]{Croft} Croft, R.~A.~C., Weinberg, D.~H., Bolte, M., et al.\ 2002, \apj, 581, 20
\bibitem[Dall'Aglio et al.(2008)]{DallAglio} Dall'Aglio, A., Wisotzki, L., \& Worseck, G.\ 2008, \aap, 491, 465
\bibitem[Dall'Aglio et al.(2009)]{DallAglio2009} Dall'Aglio, A., Wisotzki, L., \& Worseck, G.\ 2009, arXiv:0906.1484
\bibitem[Davidsen et al.(1996)]{Davidsen1996} Davidsen, A.~F., Kriss, G.~A., \& Zheng, W.\ 1996, \nat, 380, 47
\bibitem[Eisenstein \& Hut(1998)]{Eisenstein} Eisenstein, D.~J., \& Hut, P.\ 1998, \apj, 498, 137
\bibitem[Fan et al.(2006a)]{Fan} Fan, X., Carilli, C.~L., \& Keating, B.\ 2006a, \araa, 44, 415
\bibitem[Fan et al.(2006b)]{Fan_et_al} Fan, X., Strauss, M.~A., Becker, R.~H., et al.\ 2006b, \aj, 132, 117
\bibitem[Fanidakis et al.(2013)]{Fanidakis} Fanidakis, N., Maccio, A.~V., Baugh, C.~M., Lacey, C.~G., \& Frenk, C.~S.\ 2013, arXiv:1305.2199
\bibitem[Faucher-Gigu{\`e}re et al.(2008)]{Faucher} Faucher-Gigu{\`e}re, C.-A., Prochaska, J.~X., Lidz, A., Hernquist, L., \& Zaldarriaga, M.\ 2008, \apj, 681, 831
\bibitem[Fechner et al.(2006)]{Fechner2006} Fechner, C., Reimers, D., Kriss, G.~A., et al.\ 2006, \aap, 455, 91
\bibitem[Garzilli et al.(2012)]{Garzilli} Garzilli, A., Bolton, J.~S., Kim, T.-S., Leach, S., \& Viel, M.\ 2012, \mnras, 424, 1723
\bibitem[Glikman et al.(2011)]{GlikmanB} Glikman, E., Djorgovski, S.~G., Stern, D., et al.\ 2011, \apjl, 728, L26
\bibitem[Haardt \& Madau(2012)]{HM2012} Haardt, F., \& Madau, P.\ 2012, \apj, 746, 125
\bibitem[Haiman \& Hui(2001)]{Haiman} Haiman, Z., \& Hui, L.\ 2001, \apj, 547, 27
\bibitem[Hogan et al.(1997)]{Hogan1997} Hogan, C.~J., Anderson, S.~F., \& Rugers, M.~H.\ 1997, \aj, 113, 1495
\bibitem[Hui \& Gnedin(1997)]{Hui1997} Hui, L., \& Gnedin, N.~Y.\ 1997, \mnras, 292, 27
\bibitem[Jakobsen et al.(1994)]{Jakobsen1994} Jakobsen, P., Boksenberg, A., Deharveng, J.~M., et al.\ 1994, \nat, 370, 35
\bibitem[Jakobsen et al.(2003)]{Jakobsen} Jakobsen, P., Jansen, R.~A., Wagner, S., \& Reimers, D.\ 2003, \aap, 397, 891
\bibitem[Jarosik et al.(2011)]{WMAP7} Jarosik, N., Bennett, C.~L., Dunkley, J., et al.\ 2011, \apjs, 192, 14
\bibitem[Jenkins et al.(2001)]{Jenkins} Jenkins, A., Frenk, C.~S., White, S.~D.~M., et al.\ 2001, \mnras, 321, 372
\bibitem[Kim et al.(2007)]{Kim} Kim, T.-S., Bolton, J.~S., Viel, M., Haehnelt, M.~G., \& Carswell, R.~F.\ 2007, \mnras, 382, 1657
\bibitem[Komatsu et al.(2011)]{Komatsu} Komatsu, E., Smith, K.~M., Dunkley, J., et al.\ 2011, \apjs, 192, 18
\bibitem[Lidz et al.(2010)]{Lidz2010} Lidz, A., Faucher-Gigu{\`e}re, C.-A., Dall'Aglio, A., et al.\ 2010, \apj, 718, 199
\bibitem[Martini \& Weinberg(2001)]{Martini} Martini, P., \& Weinberg, D.~H.\ 2001, \apj, 547, 12
\bibitem[McDonald et al.(2001)]{McDonald2001} McDonald, P., Miralda-Escud{\'e}, J., Rauch, M., et al.\ 2001, \apj, 562, 52
\bibitem[McDonald et al.(2005)]{McDonald2005} McDonald, P., Seljak, U., Cen, R., et al.\ 2005, \apj, 635, 761
\bibitem[McDonald et al.(2006)]{McDonald2006} McDonald, P., Seljak, U., Burles, S., et al.\ 2006, \apjs, 163, 80
\bibitem[McQuinn et al.(2009)]{McQuinn} McQuinn, M., Lidz, A., Zaldarriaga, M., et al.\ 2009, \apj, 694, 842
\bibitem[Meiksin(1994)]{Meiksin1994} Meiksin, A.\ 1994, \apj, 431, 109
\bibitem[Meiksin(2009)]{Meiksin2009} Meiksin, A.~A.\ 2009, Reviews of Modern Physics, 81, 1405
\bibitem[Meiksin \& Tittley(2012)]{Meiksin} Meiksin, A., \& Tittley, E.~R.\ 2012, \mnras, 423, 7
\bibitem[Miralda-Escud{\'e} \& Rees(1994)]{Miralda1994} Miralda-Escud{\'e}, J., \& Rees, M.~J.\ 1994, \mnras, 266, 343
\bibitem[Paschos et al.(2007)]{Paschos} Paschos, P., Norman, M.~L., Bordner, J.~O., \& Harkness, R.\ 2007, arXiv:0711.1904
\bibitem[Porciani et al.(2004)]{Porciani} Porciani, C., Magliocchetti, M., \& Norberg, P.\ 2004, \mnras, 355, 1010
\bibitem[Rauch(1998)]{Rauch} Rauch, M.\ 1998, \araa, 36, 267
\bibitem[Reimers et al.(1997)]{Reimers1997} Reimers, D., Kohler, S., Wisotzki, L., et al.\ 1997, \aap, 327, 890
\bibitem[Reimers et al.(2005)]{Reimers2005} Reimers, D., Fechner, C., Hagen, H.-J., et al.\ 2005, \aap, 442, 63
\bibitem[Ricotti et al.(2000)]{Ricotti} Ricotti, M., Gnedin, N.~Y., \& Shull, J.~M.\ 2000, \apj, 534, 41
\bibitem[Rudie et al.(2012)]{Rudie} Rudie, G.~C., Steidel, C.~C., \& Pettini, M.\ 2012, \apjl, 757, L30
\bibitem[Schaye et al.(2000)]{Schaye} Schaye, J., Theuns, T., Rauch, M., Efstathiou, G., \& Sargent, W.~L.~W.\ 2000, \mnras, 318, 817
\bibitem[Shen et al.(2007)]{ShenLifetime} Shen, Y., Strauss, M.~A., Oguri, M., et al.\ 2007, \aj, 133, 2222
\bibitem[Shull \& van Steenberg(1985)]{Shull-vanSteenberg-1985} Shull, J.~M., \& van Steenberg, M.~E.\ 1985, \apj, 298, 268
\bibitem[Shull et al.(2010)]{Shull} Shull, J.~M., France, K., Danforth, C.~W., Smith, B., \& Tumlinson, J.\ 2010, \apj, 722, 1312
\bibitem[Silk \& Rees(1998)]{Silk} Silk, J., \& Rees, M.~J.\ 1998, \aap, 331, L1
\bibitem[Sokasian et al.(2002)]{Sokasian2002} Sokasian, A., Abel, T., \& Hernquist, L.\ 2002, \mnras, 332, 601
\bibitem[Sokasian et al.(2003)]{Sokasian} Sokasian, A., Abel, T., Hernquist, L., \& Springel, V.\ 2003, \mnras, 344, 607
\bibitem[Syphers et al.(2009a)]{Syphers_S} Syphers, D., Anderson, S.~F., Zheng, W., et al.\ 2009a, \apjs, 185, 20
\bibitem[Syphers et al.(2009b)]{Syphers}   Syphers, D., Anderson, S.~F., Zheng, W., et al.\ 2009b, \apj, 690, 1181
\bibitem[Syphers et al.(2012)]{Syphers2012} Syphers, D., Anderson, S.~F., Zheng, W., et al.\ 2012, \aj, 143, 100
\bibitem[Telfer et al.(2002)]{Telfer} Telfer, R.~C., Zheng, W., Kriss, G.~A., \& Davidsen, A.~F.\ 2002, \apj, 565, 773
\bibitem[Teyssier(2002)]{Teyssier} Teyssier, R.\ 2002, \aap, 385, 337 
\bibitem[Theuns et al.(2002)]{Theuns} Theuns, T., Bernardi, M., Frieman, J., et al.\ 2002, \apjl, 574, L111
\bibitem[Tittley \& Meiksin(2007)]{Tittley2007} Tittley, E.~R., \& Meiksin, A.\ 2007, \mnras, 380, 1369
\bibitem[Vanden Berk et al.(2001)]{Vanden} Vanden Berk, D.~E., Richards, G.~T., Bauer, A., et al.\ 2001, \aj, 122, 549
\bibitem[Viel et al.(2009)]{Viel} Viel, M., Bolton, J.~S., \& Haehnelt, M.~G.\ 2009, \mnras, 399, L39
\bibitem[Worseck \& Prochaska(2011)]{Worseck_Prochaska} Worseck, G., \& Prochaska, J.~X.\ 2011, \apj, 728, 23
\bibitem[Worseck et al.(2011)]{Worseck} Worseck, G., Prochaska, J.~X., McQuinn, M., et al.\ 2011, \apjl, 733, L24
\bibitem[Wyithe \& Loeb(2002)]{Wyithe} Wyithe, J.~S.~B., \& Loeb, A.\ 2002, \apj, 581, 886
\bibitem[Zaldarriaga et al.(2001)]{Zaldarriaga} Zaldarriaga, M., Hui, L., \& Tegmark, M.\ 2001, \apj, 557, 519
\bibitem[Zheng et al.(2004)]{Zheng2004} Zheng, W., Kriss, G.~A., Deharveng, J.-M., et al.\ 2004, \apj, 605, 631
\end{thebibliography}
\end{document}